\documentclass[aps,pra,amsmath,amssymb,twocolumn,preprintnumbers,superscriptaddress,10pt,longbibliography]{revtex4-2}

\usepackage[utf8]{inputenc}                                             

\usepackage[dvipsnames]{xcolor}                                                     
\usepackage{times}

\usepackage{amsmath,amssymb}                                            
\usepackage{amsthm}                                                     
\usepackage[italicdiff]{physics}                                        
\usepackage{bm}                                                         
\usepackage{tensor}                                                     
\usepackage{siunitx}                                                    
\usepackage{simplewick}                                                 

\usepackage{graphicx}                                                   
\usepackage{tikz}														
\usepackage{pgfplots}                                                   
\pgfplotsset{compat=1.14}
\usetikzlibrary{spy}

\usepackage{hyperref}                                                   

\newcommand*{\hc}{\text{h.c.}}                                          

\renewcommand*{\ket}[1]{\left| #1 \right\rangle}                        

\newcommand*{\fref}[1]{Fig.~\ref{#1}}                                   

\begin{document}


\title{Proposal for Observing Nonclassicality in Highly Excited\\ Mechanical Oscillators by Single Photon Detection}

\author{Kai Ryen Bush}
\email{Kai.Ryen@usn.no}
\affiliation{Department of Science and Industry Systems, University of South-Eastern Norway, 3616 Kongsberg, Norway}

\author{Kjetil B{\o}rkje}
\email{Kjetil.Borkje@usn.no}
\affiliation{Department of Science and Industry Systems, University of South-Eastern Norway, 3616 Kongsberg, Norway}

\date{\today}

\begin{abstract}
The preparation of pure quantum states with high degrees of macroscopicity is a central goal of ongoing experimental efforts to control quantum systems. We present a state preparation protocol which renders a mechanical oscillator with an arbitrarily large coherent amplitude in a manifestly nonclassical state. The protocol relies on coherent state preparation followed by a projective measurement of a single Raman scattered photon, making it particularly suitable for cavity optomechanics. The nonclassicality of the state is reflected by sub-Poissonian phonon statistics, which can be accessed by measuring the statistics of subsequently emitted Raman sideband photons. The proposed protocol would facilitate the observation of nonclassicality of a mechanical oscillator that moves macroscopically relative to motion at the single-phonon level.
\end{abstract}


\maketitle

In the past few decades, experimentalists have made tremendous progress moving the boundary of quantum theory's known validity to larger scales. This is motivated by fundamental questions related to, e.g., unknown decoherence mechanisms \cite{Arndt2014NatPhys} and quantum gravity \cite{Bose2017PRL,Marletto2017PRL}, but also by the prospect of new technologies taking advantage of quantum effects. The degree of macroscopicity of a quantum system can refer to physical characteristics such as mass or volume. However, given a particular system, it can also be meaningful to quantify how macroscopic its quantum state is, or how macroscopically distinct the components of a quantum superposition is \cite{Leggett1980ProgTheorPhysSuppl}. While precisely defining macroscopic quantumness is far from trivial, several attempts have been made \cite{Frowis2018RMP}.

Recent experiments in quantum optics have reported the observation of micro-macro \cite{Bruno2013NatPhys,Lvovsky2013NatPhys} and macro-macro \cite{Sychev2019Optica,Biagi2020PRL} entanglement of light, exploiting coherent displacement operations and heralded creation of single quanta (photons). The macroscopicity refers in this case to the fact that, for a bosonic mode, a coherent state $|\beta\rangle = D(\beta) |0\rangle$ and the displaced first excited state $|\beta,1\rangle = D(\beta) |1\rangle$, where $D(\beta)$ is the displacement operator and $|\beta|$ is the coherent amplitude in units of zero point fluctuations, can be distinguished by a course-grained measurement, i.e., a measurement with macroscale resolution \cite{Sekatski2014PRA}. This is possible since, even though the average Fock state occupation number of the states $|\beta\rangle$ and $|\beta,1\rangle$ only differ by 1, their number {\it distributions} differ significantly over a number range that scales with the amplitude $|\beta|$ \cite{Oliveira1990PRA}. 

Beyond purely optical systems, observing nonclassicality in large-scale {\it mechanical} oscillators is also actively being pursued. In cavity optomechanical systems \cite{Aspelmeyer2014RMP}, coherent driving of optical or microwave resonators can force mechanical oscillators into pure Gaussian states, such as the ground state \cite{Teufel2011Nature,Chan2011Nature}, the squeezed vacuum state \cite{Pirkkalainen2015PRL,Lecocq2015PRX,Wollman2015Science}, and the coherent state \cite{Wang2023PhDThesis}. Heralded single quanta (phonons) \cite{Riedinger2016Nature,Hong2017Science,Velez2019PRX} have also been realized in such systems by exploiting single-photon detection on the mechanically induced Raman sidebands of coherent optical drives, a technique demonstrated with phonons in diamond \cite{Lee2011Science,Velez2019PRX}, vibrational breathing modes in silicon nanobeams \cite{Cohen2015Nature,Riedinger2016Nature,Hong2017Science,Patel2021PRL}, flexural modes of silicon nitride membranes \cite{Galinskiy2020Optica}, elastic whispering-gallery waves in barium fluoride microresonators \cite{Enzian2021PRL,Enzian2021PRL_2}, and standing density waves in superfluid helium \cite{Patil2022PRL}. While heralded preparation of a single phonon relies on initializing the oscillator in the ground state, other state preparation proposals along these lines assumes an initial squeezed \cite{Milburn2016PRA,Shomroni2020PRA,Zhan2020PRA}, coherent \cite{Li2018PRA}, or general Gaussian \cite{Vanner2013PRL} mechanical state. Given the increased availability of both pure Gaussian state preparation and single phonon operations, it is worth investigating how ideas on macroscopicity from quantum optics could be applied to mechanical systems. 

The proposal presented here is based on initializing the mechanical oscillator in a coherent state, and subsequently realizing a {\it superposition} of phonon-addition and phonon-subtraction processes via detection of a single sideband photon, as discussed in Refs.~\cite{Vanner2013PRL,borkje2021nonclassical} and below. When starting from a coherent state $\ket{\beta}$, such a photon detection event projects the mechanical oscillator into a superposition $\ket{\psi} \propto (k_Rb+k_Bb^\dagger)\ket{\beta}$, where $b$ is the phonon annihilation operator and $k_R$ and $k_B$ are complex coefficients. Some properties of this type of state have been explored in Refs.~\cite{Moya-Cessa1995JModOpt,lee2010quantum,rahim2022enhancing}. The case $k_R = 0$, leading to a phonon-added coherent state $b^\dagger |\beta \rangle$, was discussed in the context of cavity optomechanics in Ref.~\cite{Li2018PRA}, where it was pointed out that nonclassical features, e.g., sub-Poissonian phonon statistics, vanish in the limit $|\beta| \rightarrow \infty$. 

Naively, one would also think that the state $|\psi\rangle$ more generally is not of particular interest in the high-displacement limit $|\beta| \rightarrow \infty$, since the single-phonon operations should only result in microscopic deviations from a coherent state. However, this is not necessarily true. To see this, we define $r = \beta^\ast + \beta k_R/k_B$ and write 
\begin{equation}
\label{eq:psiSPb}
    |\psi\rangle = \frac{1}{\sqrt{1 + |r|^2}} \left(r |\beta \rangle + |\beta,1\rangle  \right) ,
\end{equation}
which is a superposition of a coherent state and a displaced single-phonon Fock state. By choosing the complex coefficients $k_R, k_B$ appropriately, the state produced can be a displaced number state (if $r = 0$) or a comparably weighted superposition of macroscopically distinct states (if $|r| \sim 1$). The displaced number state is manifestly nonclassical in the sense that its Wigner distribution has a region of negativity which in principle can be verified by full quantum state tomography \cite{Enzian2021PRL,Patel2021PRL}. Another, more accessible, signature of nonclassicality is negativity of the Mandel $Q$ parameter, which reflects sub-Poissonian phonon statistics. This can be measured in the proposed setup, as described below. We will show that, for particular nonzero $r$, the Mandel $Q$ parameter of the superposition \eqref{eq:psiSPb} can be negative, also in the limit where the coherent amplitude $|\beta| \rightarrow \infty$. In this sense, the proposed protocol prepares a {\it macroscopically moving} mechanical oscillator in a manifestly nonclassical state. While motion is a relative concept, it here refers to the lab frame or any other inertial frame. We also investigate how robust the observable nonclassicality is in the more realistic case where the initial state is not exactly the ideal coherent state but has some residual thermal fluctuations. 

While we focus on cavity optomechanics below, we note that the proposed protocol can apply to other degrees of freedom that cause Raman scattering of light, e.g., magnons \cite{Kusminskiy2016PRA,Bittencourt2019PRA}. 

{\it Nonclassicality in the high-displacement limit.--} The Mandel $Q$-parameter \cite{Mandel:79} is defined as 
\begin{equation}
    Q = \frac{\expval{\Delta n^2}}{\expval{n}} - 1,
\end{equation}
where $n = b^\dagger b$ is the number operator and $\langle \Delta n^2 \rangle = \langle(n - \langle n\rangle)^2\rangle$ is the number variance.
The $Q$-parameter is the fractional deviation of the number variance from the Poisson variance $\langle n \rangle$, and thus related to the Fano factor $F$ by $Q = F-1$. 
In a classical description, $n$ represents the square of the oscillator's amplitude, which necessarily gives $Q \geq 0$. Sub-Poissonian statistics ($Q < 0$) is thus a clear signature of the nonclassical nature of the oscillator. %

For a bosonic mode in the superposition \eqref{eq:psiSPb}, we find that 
\begin{equation} \label{eq:Q_ideal_highdisp}
    \lim_{|\beta| \to \infty} Q = 2 \, \frac{1 - |r|^2 \cos (2\phi)}{(1+|r|^2)^2}
\end{equation}
in the high-displacement limit, when defining $\arg(\beta r) = \phi$. In this limit, $Q$ is minimized by the phase relation $\cos (2\phi) =  1$, with a minimal value of $Q_{\text{min}} = -1/4$ at $|r| = \sqrt{3}$. Conversely, $Q$ is maximized by $r = 0$, i.e., when $|\psi\rangle = \ket{\beta,1}$ and $\langle \Delta n^2 \rangle/\langle n \rangle = 3$ \cite{Oliveira1990PRA}, taking the value $Q_{\text{max}} = 2$.

{\it Model.--}
\begin{figure}
\includegraphics[]{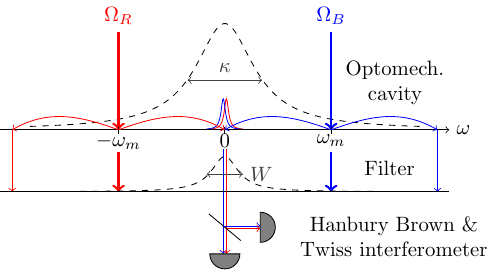}
\caption{\label{fig:Multitone_Scattering_Diagram}
       Diagram of the Raman scattering processes in the system, where $\Delta_c = 0$ and frequencies are relative to the cavity resonance frequency. Anti-Stokes photons from the red ($\Omega_R$) drive and Stokes photons from the blue ($\Omega_B$) drive pass through a filter with bandwidth $W \ll \omega_m$. The backaction from detecting a sideband photon is ambiguous, i.e., it produces a superposition of a phonon-added and a phonon-subtracted mechanical state.
       }
\end{figure}
We consider a standard optomechanical setup consisting of an optical cavity mode with (angular) resonance frequency $\omega_\mathrm{opt}$ coupled to a mechanical oscillator mode with resonance frequency $\omega_m$ via the radiation pressure interaction. As shown in Figure \ref{fig:Multitone_Scattering_Diagram}, the cavity mode is coherently driven by two lasers whose frequencies are, respectively, blue and red detuned by $\omega_m$ from their average frequency $\omega_\mathrm{ave} = \omega_\mathrm{opt} + \Delta_c$, where ideally $\Delta_c = 0$. We note that the proposed protocol is not very sensitive to nonzero $|\Delta_c| \ll \kappa$, and that the frequency spacing between the two drives can be controlled with great accuracy when the two tones originate from the same laser. In frames rotating at $\omega_\mathrm{ave}$ for the cavity mode and $\omega_m$ for the mechanical mode, the system is described by the Hamiltonian
\begin{equation}\label{eq:system_hamiltonian}
\begin{aligned}
    H(t) & = -\hbar \Delta_c a^\dagger a + \hbar g_0 a^\dagger a (e^{-i \omega_m t} b+ e^{i \omega_m t} b^\dagger) \\
    & + i\hbar \sum_{j = \{R,B\}} (a^\dagger \Omega_j(t) e^{-i\omega_j t} - a \Omega^*_j(t) e^{i\omega_j t}),
\end{aligned}
\end{equation}
where $g_0$ is the single-photon optomechanical coupling rate, $\omega_{R/B} = \mp\omega_m $ are the drive frequencies in the rotating frame, and $\Omega_{R/B}(t)$ are the corresponding drive amplitudes proportional to the square root of the power of the respective laser drives. The operators $a, \ a^\dagger$ and $b, \ b^\dagger$ are the photon and phonon ladder operators of the cavity and oscillator modes, respectively. 

We allow the drive amplitudes $\Omega_j$ to be time dependent, as the protocol presented below relies on the application of laser pulses similarly as in previous relevant experiments \cite{Riedinger2016Nature,Hong2017Science}. However, we emphasize that the experiment we propose can also be adjusted to continuous-wave operation. We present details of such a steady-state scheme in Ref.~\cite{Supplementary}, which is relevant to experiments where optical absorption is not a concern, such as with dielectric membranes \cite{Galinskiy2020Optica} or superfluid helium \cite{Patil2022PRL,Wang2023PhDThesis}.  

The system's coupling to the environment leads to dissipation, and we denote the energy damping rates of the optical and mechanical modes by $\kappa$ and $\gamma$, respectively, where typically $\gamma \ll \kappa$. We also assume that the drive frequencies obey $|\Delta_c| \ll \kappa$, which means that the ``innermost" sidebands, i.e., the upconverted (or anti-Stokes) mechanical sideband from the red detuned laser drive and the downconverted (or Stokes) sideband from the blue detuned drive depicted in Figure \ref{fig:Multitone_Scattering_Diagram}, both fall well within the cavity linewidth.
Dissipation is associated with vacuum noise from the electromagnetic environment, assuming a temperature $T$ such that $\hbar \omega_\mathrm{opt} \gg k_B T$, and vacuum and thermal noise from the mechanical environment with associated thermal occupation number $n_\mathrm{th} = 1/[\mathrm{exp}(\hbar \omega_m/k_B T) - 1]$ in the Markovian approximation \cite{Aspelmeyer2014RMP}.

We assume that light emitted from the cavity is sent through a frequency filter with bandwidth $W$ centered around the cavity resonance frequency $\omega_\mathrm{opt}$, where the bandwidth satisfies $|\Delta_c|  \ll W \ll \omega_m$.
This ensures that the frequency filter lets the two innermost mechanical sidebands through, but rejects light at the carrier frequencies of the blue and red detuned drives as well as the ``outermost" sidebands (i.e., the downconverted (upconverted) sideband from the red (blue) detuned drive). The filtered light eventually reaches single photon detectors in a Hanbury Brown and Twiss interferometer, as indicated in Figure \ref{fig:Multitone_Scattering_Diagram} and implemented in several experimental platforms \cite{Cohen2015Nature,Riedinger2016Nature,Hong2017Science,Patel2021PRL,Galinskiy2020Optica,Enzian2021PRL,Enzian2021PRL_2,Patil2022PRL}.


{\it State preparation by projective measurement.--} Let us now consider that the optomechanical system at a time $t=0$ is in the initial state $ \rho(0) = |0\rangle \langle 0 | \otimes \rho_m$, where $|0\rangle$ is the vacuum of the cavity mode and where $\rho_m$ is the inital reduced density matrix of the mechanical mode, which is ideally the coherent state $|\beta \rangle \langle \beta|$. This can for example come about from optical driving with an intensity beat note at the mechanical frequency combined with sideband cooling \cite{Li2018PRA,Bonaldi2020EurPhysJD,Wang2023PhDThesis,Supplementary}.

Starting at $t=0$, we assume that both red and blue detuned drive pulses are applied for a time $\tau_w$, and we consider flattop pulses for simplicity. The cavity field is then displaced according to $a(t) \rightarrow \bar{a}_R e^{-i\omega_R t} + \bar{a}_B e^{-i\omega_B t} + a(t)$, where $\bar{a}_{R/B} = \Omega_{R/B}/[\kappa/2 - i(\Delta_c \mp\omega_m  )]$ and $a(t)$ now describes the cavity field beyond the coherent tones. We ignore the transient buildup of the cavity field amplitudes $\bar{a}_j$ on the time scale $1/\kappa \ll \tau_w$, as the mechanical oscillator's evolution is approximately free over such short times. We apply the standard linearization procedure of only retaining terms of first order in $a(t)$ in the optomechanical interaction Hamiltonian \cite{Aspelmeyer2014RMP}. Linearization is a good approximation in the experimentally relevant regime $\kappa/g_0 \gg \max (1, \sqrt{\langle b^\dagger b \rangle}) $, i.e., when the frequency modulation due to the motion of the mechanical oscillator is small compared to the cavity linewidth.

For short pulse times $\tau_w \ll  1/[\gamma (n_\mathrm{th}+1)]$, we can neglect the mechanical mode's interaction with its intrinsic bath, and its DC response to a shift in the average radiation pressure force. We also assume that the system is in the resolved sideband regime $\kappa \ll \omega_m$ which allows the neglect of off-resonant scattering to the outermost sidebands relative to resonant scattering to the innermost sidebands. The equations of motion during the pulse can then be approximated by
\begin{align}
\label{eq:aEOM}  \dot{a} & =  -\frac{\kappa}{2} a - i \left(G_R b + G_B b^\dagger \right) + \sqrt{\kappa} a_\mathrm{in} , \\
  \dot{b} & = -i \left(G_R^\ast a + G_B a^\dagger \right) ,
\end{align}
where $G_{R/B} = g_0 \bar{a}_{R/B}$ and $a_\mathrm{in}$ is the quantum vacuum noise entering the cavity from the electromagnetic environment. We define the phases of the drive amplitudes such that $\mathrm{Im} \, (G_R G_B) = 0$, without loss of generality. Using standard input-output theory \cite{Gardiner1985PRA}, we define the total output operator as $a_\mathrm{out} = \sqrt{\kappa} a - a_\mathrm{in}$. For a narrow bandwidth signal, $a_\mathrm{out}$ is proportional to the positive frequency part of the electric field emitted from the cavity. 

In the following, we consider the limit $|G_{R/B}| \ll \kappa$, meaning that the cavity field adiabatically follows the mechanical oscillator dynamics. We define the {\it temporal} input and output modes \cite{Hofer2011PRA}
\begin{align}
  A_\mathrm{in/out} & =  \sqrt{\frac{\pm 2 {\cal G}_w}{e^{\pm 2{\cal G}_w\tau_w} - 1}} \int_0^{\tau_w} dt \, e^{\pm{\cal G}_w t} a_\mathrm{in/out}(t) , 
\end{align}
with ${\cal G}_w = (\gamma_R - \gamma_B)/2$ and $\gamma_i = 4|G_i|^2/\kappa$, which obey $[A_i,A_i^\dagger] = 1$ and are well-defined also for ${\cal G}_w \rightarrow 0$. In terms of these modes, the system's time evolution during the pulse can, in the adiabatic limit, be expressed as $A_\mathrm{out} =  U^\dagger A_\mathrm{in} U$, $b(\tau_w) = U^\dagger b(0) U$, where the evolution operator is 
\begin{align}
  U = \mathrm{exp}\left\{-i\left[A_\mathrm{in}^\dagger (k_R b(0) + k_B b^\dagger(0)) + \mathrm{h.c}\right]\right\} 
\end{align}
and we define the coefficients $k_R$ and $k_B$ by $k_R/k_B = G_R/G_B$ and $\cos(\sqrt{|k_R|^2 - |k_B|^2}) = e^{-{\cal G}_w \tau_w}$.
Note that this is well-defined for both signs of ${\cal G}_w$. This means that, in the Schr\"{o}dinger picture, the state of the system at the end of the pulse is $\rho(\tau_w) = U \rho(0) U^\dagger \equiv \sum_{i,j} |i\rangle \langle j| \otimes \rho_{m}^{(i,j)}$. For short pulses such that $\gamma_i \tau_w \ll 1$, the probability of detecting more than one photon at the cavity resonance frequency becomes negligible. In this limit, where $|k_i| \approx \sqrt{\gamma_i \tau_w}$, we find that conditioned on a single photon detection event, the mechanical state $\rho_{m,c} = \rho_{m}^{(1,1)}$ at the end of the pulse is 
\begin{align}
\label{eq:CondState}
  \rho_{m,c} = \frac{P \rho_m P^\dagger}{\mathrm{Tr}[P^\dagger P \rho_{m}]} 
\end{align}
with $P = k_R b + k_B b^\dagger$. If the initial state is a coherent state, i.e., $\rho_m = |\beta \rangle \langle \beta|$, the conditional state \eqref{eq:CondState} becomes $\rho_{m,c} = |\psi\rangle \langle \psi|$, which is the pure state defined by Equation \eqref{eq:psiSPb}. 

Defining $\lambda = |G_R/G_B|$ and $\theta = \arg(G_R G_B^\ast \beta^2)$, we can now write
$|r|^2 = |\beta|^2 [(1 - \lambda)^2 + 2 \lambda (1 + \cos \theta )]$,
where $r$ is the coefficient in Equation \eqref{eq:psiSPb}. This shows that to realize $|r| \sim 1$ (or $|r| = 0$) for a large initial amplitude $|\beta|$, we must tune the drive amplitudes such that their strength ratio $\lambda$ is close to unity and their phases (relative to the phase of the coherent state) satisfy $\theta \approx \pi$. More specifically, for $\cos(2\phi) = 1$ (where $\phi = \arg(\beta r)$ as before) or $|r| = 0$, we need to tune $\lambda$ to the optimal value $\bar{\lambda} = 1 \pm |r|/|\beta|$ and $\theta$ to the optimal $\bar{\theta} = \pi$. Physically, these conditions reflect that the coherent contributions to the two innermost sidebands must interfere destructively (for $|r| = 0$) or almost destructively (for $|r| \sim 1$) at the detectors in order for the quantum backaction from the single-photon detection event to have a large impact on the mechanical state. We note that the phase matching requirement can most easily be met if the initial coherent state is also produced optically by the application of an intensity beat note. Assuming two beams of equal amplitude, a small devation of $\lambda$ from unity can simply be implemented by slightly detuning both beams by a frequency $\approx \pm (\kappa^2/(4 \omega_m) + \omega_m) \sqrt{|r|/|\beta|}$ from the mechanical sidebands, as long as the beams are locked more accurately than this to variations in the cavity resonance frequency.

An alternative method of preparing the state \eqref{eq:psiSPb} is by displacing the superposition $(r|0\rangle + |1\rangle)$ of number states by the application of a coherently oscillating mechanical force with negligible noise on a time scale much faster than the oscillator's decoherence time. However, for $r \neq 0$, this simply corresponds to a different order of the operations in the proposed protocol which is more sensitive to decoherence. It is also possible to prepare the state by replacing the phonon annihilation term $b |\beta\rangle$ by an identity operation, i.e., $|\psi \rangle \propto (\mu \mathbb{I} + b^\dagger)|\beta\rangle$ with $ \mu = r - \beta^\ast$, which in principle can be implemented by adding a weak coherent tone at the cavity resonance frequency to the signal reaching the photodetectors \cite{Vanner2013PRL}. This also has no obvious advantage, as it would require the same level of control over the relative power and the phases of pulses at different frequencies. 

\begin{figure}
\includegraphics[scale=0.95]{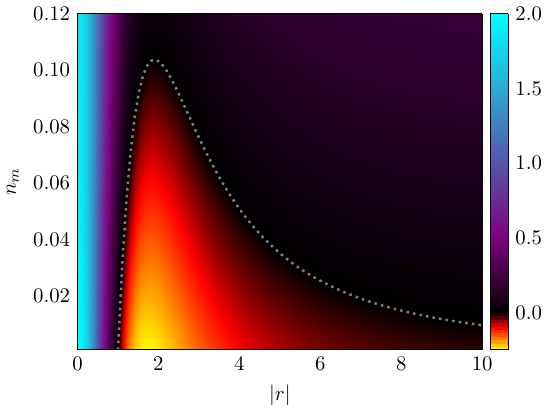}
\noindent\caption{\label{fig:mandelQ_r_n}
       The Mandel $Q$ parameter \eqref{eq:QbcDisplacedThermal} for the conditional state \eqref{eq:CondState} when the inital state $\rho_m$ is a displaced, thermal state, in the high displacement limit $\abs{\beta}\gg 1$, as a function of $|r|$ and the initial average phonon number $n_m$. We have assumed the ideal phase relation $\cos (2\phi)=1$. The dotted line separates the regions of positive and negative $Q$.
       }
\end{figure}

In a realistic experimental setting, the initial state will not be the ideal coherent state, but more likely a displaced, thermal state $\rho_m = D(\beta) \rho_\mathrm{th} D^\dagger(\beta)$, where
\begin{align}
\label{eq:ThermalState}
 \rho_\mathrm{th} = \sum_{n = 0}^\infty \frac{n_m^n}{(n_m + 1)^{n + 1}} |n \rangle \langle n |  
\end{align}
is a thermal state with average phonon occupation number $n_m$. Given this initial state, we denote the Mandel $Q$ parameter of the mechanical mode in the conditional state \eqref{eq:CondState} by $Q_{b,c}$, whose general expression can be found in the Ref.~\cite{Supplementary}. Here, we are interested in the high-displacement limit
\begin{align}
\label{eq:QbcDisplacedThermal}
\lim_{|\beta| \to \infty} Q_{b,c} = 2 \left(\frac{1+2n_m-|r|^2 \cos(2\phi)}{(1+2n_m+|r|^2)^2} + n_m \right) .
\end{align}
This shows that $Q_{b,c}$ is minimized by choosing parameters such that $\cos (2\phi) = 1$ and $|r| = \sqrt{3 (1 +2n_m)}$, in which case we find that $Q_{b,c} < 0$ requires the initial thermal occupation $n_m < 0.10$. Figure \ref{fig:mandelQ_r_n} shows the high-displacement value of $Q_{b,c}$ in Equation \eqref{eq:QbcDisplacedThermal} as a function of $|r|$ and $n_m$, given the ideal phase relation $\cos(2\phi) = 1$. We see that $Q_{b,c}$ can be negative also for nonideal values of $|r|$, but then with stricter limits on the maximal thermal occupation $n_m$. While the proposed protocol has a very low tolerance for thermal noise, we note that this type of experiment on high-frequency silicon nanobeams in a cryogenic environment have reported $n_m < 0.025$ \cite{Riedinger2016Nature}, albeit in the absence of coherent oscillations.

Experimental inaccuracies in the amplitude ratio or the relative phase of the two drives, i.e., in $\lambda$ and $\theta$, will lead to deviations from the optimal choices of $|r|$ and $\phi$ that minimizes $Q_{b,c}$. To second order in $\Delta \lambda = \lambda - \bar{\lambda}$ and $\Delta \theta = \theta - \bar{\theta}$, we find that the Mandel $Q$ parameter deviates from its minimal value according to
\begin{align}
\label{eq:QbcCorrections}
\Delta Q_{b,c} = \frac{|\beta|^2}{4(1 + 2n_m)^2} \left(\frac{3}{4 } \Delta \lambda^2 + \Delta \theta^2\right) ,
\end{align}
again, in the limit $|\beta| \rightarrow \infty$. The observation of sub-Poissonian statistics thus requires that the amplitudes and phases of the two beams can be controlled such that $\Delta \lambda, \Delta \theta \ll 1/|\beta|$, becoming increasingly demanding for increasing $|\beta|$. However, implementation of the protocol with an initial average phonon number of $|\beta|^2 = 100$ (as an example) would only require control of relative amplitude and phase (in radians) at the percent level. This seems well within present capabilites if the two beams originate from the same laser by acousto- or electrooptic modulation. 

{\it Characterization of the conditional state.--} The phonon statistics of the conditional state \eqref{eq:CondState} can be accessed by applying another pulse, this time red detuned only, starting at time $t_r$ and with duration $\tau_r$, and detecting anti-Stokes photons upconverted to the cavity resonance frequency \cite{Riedinger2016Nature,Hong2017Science,Li2018PRA}. According to the above discussion, the corresponding temporal input and output modes will then be related by
\begin{align}
\label{eq:AoutRead}
  A_\mathrm{out} & =  e^{- {\cal G}_r \tau_r} A_\mathrm{in} - i \sqrt{1 - e^{-2 {\cal G}_r \tau_r}}  \, b(t_r) , 
\end{align}
with ${\cal G}_r = 2|G_R|^2/\kappa$. This shows that the number statistics of resonant sideband photons generated by the pulse relate to the phonon statistics, since the input vacuum noise $A_\mathrm{in}$ to the cavity does not contribute to normal ordered correlation functions of $A_\mathrm{out}$. If the photodetector dead times exceed the pulse duration, as in Refs.~\cite{Riedinger2016Nature,Hong2017Science}, the pulse will give rise to either $ n = 0$, 1, or 2 detection events with respective probabilities $p_n$ (nontrivially) related to the phonon number probability distribution. In the limit $\eta (1 - e^{-2 {\cal G}_r \tau_r}) \ll 1$, where $\eta$ is the fraction of cavity sideband photons that are detected when accounting for other cavity decay channels, transmission loss, and nonunit detector efficiency, we then have the simplified relation $\eta (1 - e^{-2 {\cal G}_r \tau_r}) Q_{b} \approx 4 p_2/p_1 - p_1$ when assuming a 50/50 beam splitter in the Hanbury Brown and Twiss interferometer. This relation holds more generally for $\tau_r \rightarrow \tau_\mathrm{bin}$ when $p_n$ is the probability of $n$ events within a chosen bin time $\tau_\mathrm{bin}$, as long as the probability of a photodetection event within $\tau_\mathrm{bin}$ is small.

{\it Concluding remarks.--} We have presented a protocol for preparing a mechanical oscillator with macroscopically large coherent amplitude, compared to motion at the single-phonon level, in a manifestly nonclassical state. The protocol based on laser pulses is, for instance, relevant to optomechanical systems involving vibrational modes in silicon nanobeams with high frequencies ($\omega_m/2\pi \sim 5$ GHz) and low thermal occupation ($n_m \ll 1$), where detection of single Raman sideband photons have already been used as a tool for observing nonclassicality \cite{Riedinger2016Nature,Hong2017Science}, provided that the vibrational mode can be initialized close to a coherent state.

The proposed experiment can, however, also be performed with steady-state laser driving in systems where optical absorption is not a concern. As an example, we consider fiber cavity optomechanics with acoustic waves in superfluid helium \cite{Patil2022PRL,Wang2023PhDThesis}. A displaced, thermal state with amplitude $|\beta| \approx 100$ and thermal occupation $n_m \sim 1$ was recently reported in this system \cite{Wang2023PhDThesis}, and sideband cooling to smaller $n_m$ is expected in future implementations \cite{Wang2023PhDThesis}. With a mechanical resonance frequency $\omega_m/2\pi \sim  300$ MHz, a cavity linewidth $\kappa \sim 50$ MHz, and a filter bandwidth $W \sim 1$ MHz, it should be feasible to {\it separately} detect the sideband photons used for state preparation and those used for read-out. As shown in Ref.~\cite{Supplementary}, by applying five continuous drive tones simultaneously, the system can, on average, reside in a steady, displaced, thermal state, while the phonon statistics conditioned on the desired projective measurement is simultaneously collected.

Finally, we note that the central idea of the protocol can also be exploited in order to prepare micro-macro and macro-macro entanglement of mechanical oscillators, following ideas from quantum optics \cite{Biagi2020PRL}.

We acknowledge useful discussions with Jack Harris.

\clearpage
\vspace{1cm}
\begin{center}
{\Large\bfseries Supplementary Material}
\end{center}

\section{A general Mandel $Q$ parameter}

\begin{figure}
\noindent\includegraphics[scale=0.95]{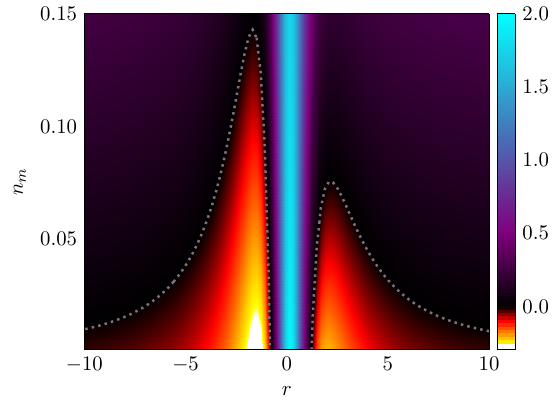}
\noindent\includegraphics[scale=0.95]{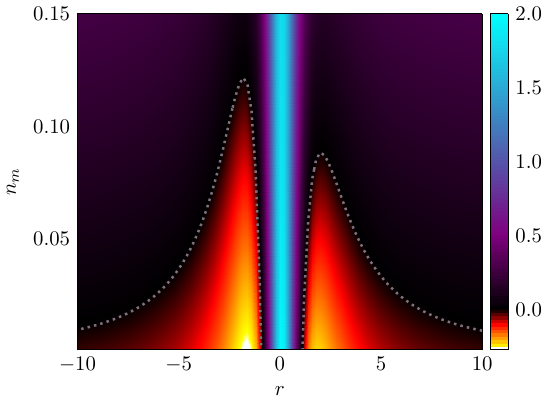}
\noindent\includegraphics[scale=0.95]{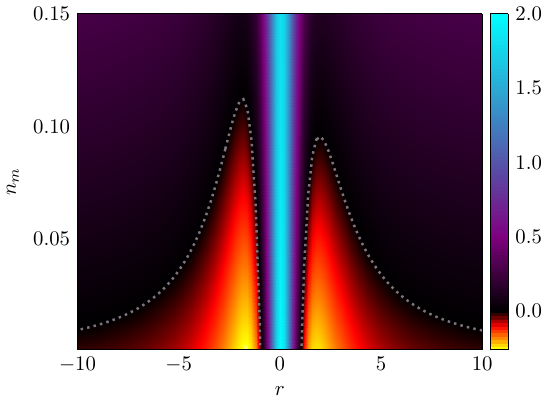}
\noindent\caption{\label{fig:mandelQ_r_n_b}
       The conditioned Mandel $Q$ parameter \eqref{eq:mandel_q_conditioned_full} evaluated for $\abs{\beta}=\{ 5, 10, 20 \}$ (from the top) along the optimizing axis $\cos(2\phi)=1$, where negative values of $r$ represent the $\cos (\phi) = -1$ part of the solution. The dotted lines delineate the negative (nonclassical) region boundaries. Note that for lower $\abs{\beta}$ the nonclassical region stretches to higher $n_m$ when $r<0$, and going as low as $Q_{b,c} \sim -0.3$ when $\abs{\beta}=5$ and $n_m=0$.
    }
\end{figure}

Here we provide a full expression for the Mandel $Q$ parameter for the projected state (9) in the main article, with $r$-, $n_m$- and $\beta$-dependence. Using $b = \beta + b_n$, we can express it in terms of noise correlators as
\begin{equation}\label{eq:mandel_q_conditioned_full}
\begin{aligned}
    Q_{b,c} = &\frac{1}{\abs{\beta}^2+\langle b_n^\dagger b_n\rangle_c+2\Re{\beta^* \expval{b_n}_c}}  \\
    &\ \ \ \ \ \ \ \times \bigg( 2\abs{\beta}^2\expval{b_n^\dagger b_n}_c +2\Re{\beta^{*2}\expval{b_n b_n}_c} \\
    &\ \ \ \ \ \ \ \ \ \ \ +4\Re{\beta^*\expval{b_n^\dagger b_n b_n}_c} +\expval{b_n^\dagger b_n^\dagger b_n b_n}_c \\
    &\ \ \ \ \ \ \ \ \ \ \ \ \ \ \ \ \ \ \ -\left(\expval{b_n^\dagger b_n}_c+2\Re{\beta^* \expval{b_n}_c}\right)^2\bigg).
\end{aligned}
\end{equation}
The conditioned expectation values
\begin{equation}
    \expval{O}_c = \frac{\expval{(k_R^*b^\dagger+k_B^* b)O(k_Rb+k_B b^\dagger)}}{\expval{(k_R^*b^\dagger+k_B^* b)(k_Rb+k_B b^\dagger)}}
\end{equation}
can be expressed in terms of thermal noise correlators $\expval{b_n^\dagger b_n}=n_m$. Using that $k_R=k_B(r-\beta^*)/\beta$ we can write them as
\begin{subequations}
\begin{align}
    \expval{b_n^\dagger b_n^\dagger b_n b_n}_c &= 2n_m\frac{3Dn_m+\abs{\beta}^2(1+2n_m(1-\abs{r}^2))}{D}, \\
    \expval{b_n^\dagger b_n b_n}_c &= 2n_m \frac{n_m r\beta(r^*-\beta)+r^*\abs{\beta}^2(1+n_m)}{D}, \\
    \expval{b_n^\dagger b_n}_c &= \frac{2Dn_m+\abs{\beta}^2(1+n_m(1-\abs{r}^2))}{D}, \\
    \expval{b_n b_n}_c &= \frac{2n_m(1+n_m)\beta(r^*-\beta)}{D}, \\
    \expval{b_n}_c &= \frac{n_m r\beta(r^*-\beta)+r^*\abs{\beta}^2(1+n_m)}{D},
\end{align}
\end{subequations}
where
\begin{equation}
    D = \abs{\beta}^2(1+\abs{r}^2)+(2\abs{\beta}^2+\abs{r}^2-2\abs{r}\abs{\beta}\cos(\phi))n_m.
\end{equation}
In \fref{fig:mandelQ_r_n_b}, $Q_{b,c}$ is plotted as a function of $r$ and $n_m$ for $\abs{\beta} \in \{ 5, 10, 20 \}$.

\section{Optomechanical coherent state preparation}
As shown in the main article, negativity of the Mandel $Q$ parameter in the high-displacement limit requires that the mechanical oscillator is initially in a displaced thermal state with a low thermal occupation number $n_m<0.10$. Preparing a mechanical oscillator in a near-coherent state is possible by leveraging built-in optomechanical effects 
in the regime $\gamma \ll \kappa$. This motivates a steady state solution of a linearized, adiabatically coupled optomechanical system driven by several drives of arbitrary frequencies.

\subsection{Steady-state solution of the linearized quantum Langevin equations}
We consider a standard optomechanical setup consisting of an optical cavity mode with resonant frequency $\omega_\mathrm{opt}$ coupled to a mechanical oscillator mode with resonant frequency $\omega_m$ via the radiation pressure coupling $g_0$. The system is driven by some arbitrary number of coherent lasers $\{ \Omega_j\}$. In the frame corotating with the cavity resonance frequency this system is described by the Hamiltonian
\begin{equation}
\begin{aligned}
    H = \hbar \omega_m b^\dagger b + \hbar g_0 a^\dagger a (b+b^\dagger)+ i\hbar \sum_j a^\dagger \Omega_je^{-i\omega_j t} +\hc
\end{aligned}
\end{equation}
where $\{ \omega_j\}$ are the frequencies of the drive lasers relative to the cavity resonance, and $a, a^\dagger$ and $b, b^\dagger$ are the ladder operators of the optical and mechanical modes respectively. Including coupling to optical and mechanical thermal baths, it obeys open system dynamics governed by the quantum Langevin equations
\begin{subequations} \label{eq:langevin_eqs_full}
\begin{align}
    \dot{a} &= -\frac{\kappa}{2}a -ig_0a(b+b^\dagger) +\sum_j \Omega_j e^{-i\omega_j t}+\sqrt{\kappa}a_{\text{in}}, \label{eq:langevin_eq_a_full}\\
    \dot{b} &= -\left(\frac{\gamma}{2}+i\omega_m\right)b-ig_0a^\dagger a+\sqrt{\gamma}b_{\text{in}} \label{eq:langevin_eq_b_full}.
\end{align}
\end{subequations}
$\kappa$ and $\gamma$ are the optical and mechanical damping rates, and $a_\mathrm{in}(t)$ and $\ b_\mathrm{in}(t)$ are the usual bosonic input noise operators for the cavity and mechanical oscillator, satisfying
\begin{align}
    \langle a_{\text{in}}(t)a_{\text{in}}^\dagger(t^\prime) \rangle &= \delta(t-t^\prime), \\
    \langle b_{\text{in}}(t)b_{\text{in}}^\dagger(t^\prime)\rangle &= (n_\mathrm{th}+1)\delta(t-t^\prime), \\
    \expval{a_{\text{in}}(t)a_{\text{in}}(t^\prime)}&=\expval{b_{\text{in}}(t)b_{\text{in}}(t^\prime)}=0
\end{align}
and $[a_{\text{in}}(t),a_{\text{in}}^\dagger(t^\prime)]=[b_{\text{in}}(t),b_{\text{in}}^\dagger(t^\prime)]=\delta(t-t^\prime)$.
The bare thermal occupation number $n_\mathrm{th} = 1/[\mathrm{exp}(\hbar \omega_m/k_B T) - 1]$ is the average number of phonons of the mechanical oscillator when in thermal equilibrium with its environment at temperature $T$, not to be confused with the effective thermal occupation number $n_m$. The optical cavity's thermal occupation number is taken to be zero, as we assume $\hbar \omega_\mathrm{opt} \gg k_B T$.
 
The usual linearization procedure is applied by splitting the ladder operators into coherent amplitudes and quantum noise terms $a(t)\to \bar{a}(t) + a_n(t), \ b(t)\to \bar{b}_c+\bar{b}(t)+ b_n(t)$, allowing us to solve to each order in the operators independently. $\bar b_c$ represents a constant shift in the equilibrium displacement of the mechanical oscillator under continuous driving. The optical coherent terms can be solved perturbatively around the non-interacting solution
\begin{equation}
    \bar a_0(t) = \sum_j \bar a_j(t) = \sum_j \frac{\Omega_j e^{-i\omega_j t}}{\frac{\kappa}{2}-i\omega_j}
\end{equation}
to arrive at
\begin{equation} \label{eq:coherent_solution_a}
    \bar{a}(t) = \bar a_0(t) +k_R(t) \bar b(t) + k_B(t) \bar b^*(t),
\end{equation}
where we have used the adiabatic approximation
\begin{equation}
\begin{aligned}
\int_{-\infty}^t e^{\frac{\kappa}{2} \tau} \bar a_0(\tau) b(\tau) \dd \tau \approx b(t) e^{i\tilde\omega_m t} \int_{-\infty}^t e^{\frac{\kappa}{2} \tau-i\tilde\omega_m \tau} \bar a_0(\tau) \dd \tau
\end{aligned}
\end{equation}
to separate out the mechanical motion (note that the mechanical oscillator is assumed to oscillate at the renormalized mechanical frequency $\tilde\omega_m$, which will be defined later). This approximation is valid when the optical cavity "forgets" the mechanical oscillator's previous states at a much faster rate than it changes, which is the case when $\gamma, g_0 \ll \kappa$. The coefficients in \eqref{eq:coherent_solution_a} are given by
\begin{subequations} \label{eq:coeffs_exact}
\begin{align}
    k_R(t) &= -ig_0\sum_j \frac{\bar{a}_j(t)}{\kappa/2-i(\omega_j+\tilde\omega_m)}, \\
    k_B(t) &= -ig_0\sum_j \frac{\bar{a}_j(t)}{\kappa/2-i(\omega_j-\tilde\omega_m)},
\end{align}
\end{subequations}
representing the magnitudes of anti-Stokes and Stokes processes respectively. The cavity frequency shift $g_0(\bar b_c+ \bar b_c^*)$ due to the shift in mechanical equilibrium is absorbed into the bare optical frequency $\omega_\mathrm{opt}$ and consequently the rotating frame, but note that changing the drive configurations may also change this shift.

Expanding the coherent mechanical equation of motion with \eqref{eq:coherent_solution_a} and neglecting second order terms in $\bar b_c+\bar b(t)$ as well as counter-rotating terms $\propto \bar b_c^* + \bar b^*(t)$ gives the solution
\begin{equation} \label{eq:coherent_solution_b}
    \bar{b}(t) = -ig_0\sum_{j\neq k}\frac{\bar{a}_j(t)\bar{a}_k^*(t)}{\tilde \gamma/2+i\tilde \omega_m-i(\omega_j-\omega_k)},
\end{equation}
with the remaining terms giving the static displacement
\begin{equation} \label{eq:coherent_solution_b_c}
    \bar{b}_c = -ig_0\sum_{j}\frac{\abs{\bar{a}_j(t)}^2}{\tilde \gamma/2+i\tilde \omega_m},
\end{equation}
which can be neglected for the remainder of the discussion.
Similarly to the optical resonance shift, the mechanical renormalizations $\tilde \gamma - \gamma =-2\Im{\Sigma}$, $\tilde \omega_m - \omega_m = \Re{\Sigma}$ are given by the self-energy
\begin{equation} \label{eq:renorm_mech}
    \Sigma = g_0 \lim_{T\to \infty}\frac{1}{2T}\int_{-T}^T (\bar a_0^*(t)k_R(t)+\bar a_0(t)k_B^*(t)) \dd t
\end{equation}
where averaging cancels out the oscillating terms and leaves only a constant contribution. 

The remaining terms which are linear in the noise operators form the linearized quantum optomechanical Langevin equations
\begin{subequations}
\begin{align}
    \dot a_n &\approx -\frac{\kappa}{2}a_n-ig_0\bar a_0(t)(b_n+b_n^\dagger)+\sqrt{\kappa}a_{\text{in}}, \label{eq:a_n_linearized_lang_eq}\\
    \dot b_n &\approx -\left(\frac{\gamma}{2}+i\omega_m\right)b_n-ig_0(\bar a_0^*(t)a_n+\bar a_0(t) a_n^\dagger)+\sqrt{\gamma}b_{\text{in}}. \label{eq:b_n_linearized_lang_eq}
\end{align}
\end{subequations}
These can be solved similarly to the interacting coherent terms. The cavity noise solution is solved adiabatically as
\begin{equation} \label{eq:a_n_solution}
    a_n(t) = k_R(t) b_n(t) + k_B(t) b_n^\dagger(t)+ \zeta(t)
\end{equation}
where we define the cavity noise operator
\begin{equation} \label{eq:zeta}
    \zeta(t) = \sqrt{\kappa}\int_{-\infty}^t e^{-\frac{\kappa}{2}(t-\tau)}a_{\text{in}}(\tau) \dd \tau.
\end{equation}
The mechanical noise equation of motion in the rotating wave approximation becomes
\begin{equation}
\begin{aligned}
    \dot b_n = -\left(\frac{\tilde \gamma}{2}+i \tilde \omega_m\right)b_n-ig_0(\bar a_0(t)\zeta^\dagger(t)+&\bar a_0^*(t)\zeta(t))+\sqrt{\gamma}b_{\text{in}},
\end{aligned}
\end{equation}
with the solution
\begin{equation} \label{eq:b_n_solution}
\begin{aligned}
    b_n&(t) = \int_{-\infty}^t e^{-(\frac{\tilde \gamma}{2}+i\tilde \omega_m)(t-\tau)}\sqrt{\gamma}b_{\text{in}}(\tau)\dd \tau \\
    &-ig_0\int_{-\infty}^t e^{-(\frac{\tilde \gamma}{2}+i\tilde \omega_m)(t-\tau)}(\bar a_0(\tau)\zeta^\dagger(\tau)+\bar a_0^*(\tau)\zeta(\tau))\dd \tau.
\end{aligned}
\end{equation}
\eqref{eq:coherent_solution_a}, \eqref{eq:a_n_solution} and \eqref{eq:zeta}, \eqref{eq:coherent_solution_b} and \eqref{eq:b_n_solution} form the complete linearized, adiabatic multi-tone solutions used in the remainder of this material.

\subsection{Optomechanical cooling and displacement}
We can investigate combined sideband cooling and beat note displacement using the multi-tone steady state solution obtained above. In principle the drives can be applied to a separate optical mode (of the same or a different cavity), or to the same mode as the measurement drives if the detector filter linewidth $W$ is much smaller than $\tilde\omega_m$, such that the measured sidebands can be isolated in the output signal. Sideband cooling is as usual achieved via a red-detuned drive $\Omega_\mathrm{cool}$ favoring anti-Stokes scattering which reduce the number of mechanical quanta. Displacement is achieved by applying drives $\Omega_{d\pm}$ to the half-sideband frequencies $\omega_{d\pm} = \pm\tilde\omega_m/2$. According to \eqref{eq:coherent_solution_b}, this induces a coherent mechanical oscillation
\begin{equation} \label{eq:approx_disp}
\begin{aligned}
    \bar b(t) \approx -i\frac{8g_0\Omega_{d+}\Omega_{d-}^*}{\tilde\gamma (\kappa- i\tilde\omega_m)^2}e^{-i\tilde\omega_m t} \equiv \beta e^{-i\tilde\omega_m t}.
\end{aligned}
\end{equation}
If the temperature of the mechanical oscillator were approximately zero it would therefore occupy a coherent steady state $D(\beta)\ket{0}$. Note that any combination of drives such that $\omega_{d+}-\omega_{d-}=\tilde\omega_m$ will produce this effect, but placing them symmetrically about the optical resonance cancels their contributions to the renormalization \eqref{eq:renorm_mech} when $\abs{\Omega_{d-}}=\abs{\Omega_{d+}}$. This method can be applied continuously to produce a steady state displaced thermal state, or switched off after the steady state has been reached, with the displacement persisting for a time of the order $1/\tilde\gamma$.

In the following we establish the effective thermal properties of the mechanical oscillator under multi-tone driving, and derive the maximal possible displacement $\abs{\beta}$ reachable while staying below a given effective thermal occupation number
\begin{equation}\label{eq:nm}
    n_m \equiv \expval{b_n^\dagger b_n}.
\end{equation}

The magnitude of the mechanical renormalizations is determined by the self-energy \eqref{eq:renorm_mech}. By discarding the mechanically off-resonant terms we can express this as a sum of single-tone contributions
\begin{equation}
\begin{aligned}
    \Sigma & \approx -ig_0^2\sum_{j} \frac{\abs{\Omega_j}^2}{\frac{\kappa^2}{4}+\omega_j^2} \\
    & \ \ \ \times \bigg(\frac{1}{\frac{\kappa}{2}-i(\tilde\omega_m+\omega_j)}-\frac{1}{\frac{\kappa}{2}-i(\tilde\omega_m-\omega_j)} \bigg).
\end{aligned}
\end{equation}
The summand is antisymmetric in $\omega_j$, meaning that symmetric drives of equal magnitude will cancel each other. Since $\abs{\Omega_R}\to \abs{\Omega_B}$ in the high displacement limit and we can freely pick displacement drives that satisfy $\abs{\Omega_{d-}}=\abs {\Omega_{d+}}$, only the cooling drive contributes to the renormalization, by the amount
\begin{equation} \label{eq:renorm_coolingtone}
\begin{aligned}
    \Sigma_\mathrm{cool} \approx \frac{-4\tilde\omega_m g_0^2\abs{\Omega_\mathrm{cool}}^2}{\kappa(\frac{\kappa^2}{4}+\tilde\omega_m^2)(\frac{\kappa}{2}-2i\tilde\omega_m)}.
\end{aligned}
\end{equation}
This gives an effective mechanical damping rate
\begin{equation} \label{eq:eff_damping}
    \tilde \gamma \approx \gamma + \frac{16 \tilde\omega_m^2 g_0^2 \abs{\Omega_\mathrm{cool}}^2}{\kappa(\frac{\kappa^2}{4}+\tilde\omega_m^2)(\frac{\kappa^2}{4}+4\tilde\omega_m^2)}
\end{equation}
which may be much larger than $\gamma$, and an effective mechanical frequency
\begin{equation} \label{eq:eff_mech_freq}
    \tilde \omega_m \approx \omega_m - \frac{2 \tilde\omega_m g_0^2 \abs{\Omega_\mathrm{cool}}^2}{(\frac{\kappa^2}{4}+\tilde\omega_m^2)(\frac{\kappa^2}{4}+4\tilde\omega_m^2)}.
\end{equation}
This shift can be neglected in our regime since $\abs{\tilde \omega_m-\omega_m} \sim \abs{\tilde \gamma-\gamma} \ll \omega_m$, but the right hand side can otherwise be solved perturbatively in $\tilde \omega_m - \omega_m$ to higher orders.

The effective mechanical damping is not the only way by which coherent driving alters the thermal properties of the optomechanical system. From \eqref{eq:b_n_solution} we find, up to mechanical resonance and using $\tilde \gamma \ll \kappa$, that the effective thermal occupation number \eqref{eq:nm} can be written in terms of the bare occupation number $n_\mathrm{th.}$ as
\begin{subequations}
\begin{equation}\label{eq:effective_n_m}
    n_m = \frac{\gamma}{\tilde \gamma}n_\mathrm{th.}+n_o,
\end{equation}
where we can write the purely optical contribution to the thermal occupation number as
{\small\begin{equation} \label{eq:optical_n}
    n_o \approx \sum_{j} \frac{g_0^2 \kappa \abs{\bar{a}_j(t)}^2}{\tilde \gamma(\frac{\kappa^2}{4}+(\tilde\omega_m+\omega_j)^2)}.
\end{equation}}
\end{subequations}
Inserting \eqref{eq:eff_damping} into this expression and taking the limit $\tilde \gamma \gg \gamma$, $\tilde \omega_m \gg \kappa$, we recover for a single red-detuned drive the known resolved sideband cooling lower bound $n_o \to (\kappa/4\tilde\omega_m)^2$. For two displacement drives of equal magnitude $\abs{\Omega_d}$ at $\{-\tilde\omega_m/2, \tilde\omega_m/2 \}$, we obtain the effective thermal contribution
\begin{align}
    n_o\approx \frac{16g_0^2\kappa \abs{\Omega_\mathrm{d}}^2}{\tilde \gamma(\kappa^2+\tilde\omega_m^2)}\left(\frac{1}{\kappa^2+\tilde\omega_m^2} +  \frac{1}{\kappa^2+9\tilde\omega_m^2} \right),
\end{align}
assuming this is the dominant contribution to $n_o$. According to \eqref{eq:coherent_solution_b}, the threshold $n_o < \epsilon$ therefore limits the amplitude to
\begin{equation}
     \abs{\beta} < \frac{(\kappa^2+\tilde\omega_m^2)(\kappa^2+9\tilde\omega_m^2)}{4 g_0 \kappa(\kappa^2+5\tilde\omega_m^2)}\epsilon,
\end{equation}
which in the resolved sideband limit becomes $|\beta| < 9\tilde{\omega}_m^2/(20 \kappa g_0) \epsilon$.

\section{Measuring in steady state}

\begin{figure}
\includegraphics[scale=0.95]{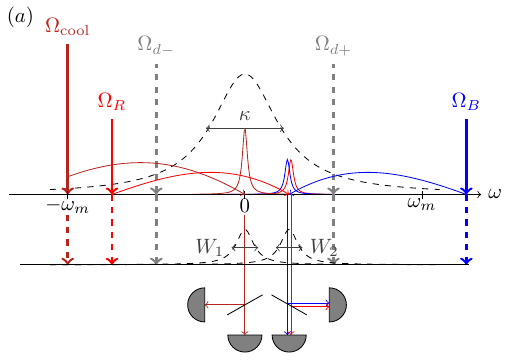}
\includegraphics[scale=0.95]{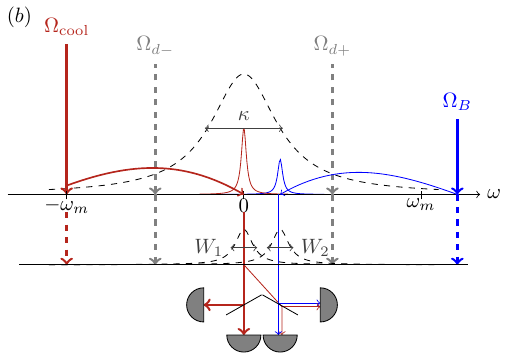}
\caption{\label{fig:Multitone_Steadystate_Scattering_Diagram}
       Schematic of steady-state operation with multitone driving. Rather than varying the drive strengths sequentially (i.e., pulsed operation), sufficiently narrow filters allow different Stokes and anti-Stokes sideband compositions to be measured simultaneously in the steady state. $(a)$ shows a five-drive configuration with optomechanical cooling and displacement. Isolating the overlapping projective measurement sidebands from the cooling/readout sideband applies in the limit $W \ll |\Delta_\mathrm{proj.}|$. $(b)$ shows an alternative four-drive configuration where a strong cooling sideband may be used as an anti-Stokes photon source for both read and write measurements simultaneously by passing through both filters.
       }
\end{figure}

Our protocol relies on two separate modes of measurement. The first is to make a projective single photon measurement on overlapping Stokes and anti-Stokes sidebands, producing the mechanical state given by Equation (9) in the main article, and the other is to measure only the anti-Stokes sideband to recover the phonon statistics of the mechanical oscillator. In the main article a conceptually simple realization of these modes of measurement is proposed where they are confined discretely to different times in a write- and read-pulse formulation. However, sufficiently narrow optical filters can eliminate the need for sequential operation by resolving the processes in the frequency domain rather than in time. The filtered measurements ideally only depend on the coefficients $k_R$ and $k_B$, and the sidebands can therefore in principle be shifted an amount less than $\sim\kappa/2$ off the cavity resonance without any loss of generality. Shifting the measurement drives $\Omega_R$, $\Omega_B$ by a frequency $\Delta_\mathrm{proj.}$ each, a separate drive $\Omega_\mathrm{cool}$ can be added to the system at $-\omega_m$, which may be used for combined phonon measurement and cooling. Two optical filters centered at $\Delta_\mathrm{proj.}$ and optical resonance respectively, with transmission linewidths $W \ll \abs{\Delta_\mathrm{proj.}}$ can then be used to alter the transmission spectra of each branch of a beamsplitter, producing two output modes proportional to
\begin{subequations} \label{eq:filtered_multitone_signals}
    \begin{align}
        a_\mathrm{proj.}(t) &\approx k_R(t)b(t)+k_B(t)b^\dagger(t)+\zeta_\mathrm{proj.}(t),\label{eq:filtered_multitone_signals_proj} \\
        a_\mathrm{cool}(t) &\approx k_\mathrm{cool}(t)b(t)+\zeta_\mathrm{cool}(t)
    \end{align}
\end{subequations}
with different sideband compositions. The filtered optical noise terms
\begin{equation}
    \zeta_f(t) = \frac{W}{2}\int_{-\infty}^\infty e^{-i\omega t} \frac{1}{W/2-i(\omega-\Delta_f)}\zeta[\omega] \dd \omega
\end{equation}
does not contribute to the correlation functions of interest. The two output modes can be measured simultaneously by independent detectors, as indicated in \fref{fig:Multitone_Steadystate_Scattering_Diagram}. A single photon detection from the signal $a_\mathrm{proj.}(t)$ projects the mechanical oscillator onto the target state, altering the phonon statistics measured from $a_\mathrm{cool}(t) \propto b(t)$.

Having established the possibility for cooling, displacement and both modes of measurement in steady state, the system can be driven continuously with up to five simultaneous drives:
\begin{itemize}
    \item Two displacement drives $\Omega_{d\pm}$ tuned to $\omega_{d\pm}=\pm\omega_m/2$.
    \item A combined readout and cooling drive $\Omega_\mathrm{cool}$ for phonon statistics measurement tuned to $\omega_\mathrm{cool}=-\omega_m$.
    \item Two measurement drives $\Omega_R$, $\Omega_B$ for projective measurement tuned to $\omega_{B/R}=\pm\omega_m+\Delta_\mathrm{proj.}$.
\end{itemize}
Although this should work in principle, with several drives and their sidebands occupying the same cavity mode there may arise issues in isolating each measurement channel. In particular, a frequency separation between drives and sidebands much larger than $W$ is not necessarily sufficient to isolate frequency bands when one or more drive magnitudes are very large, e.g. with sideband cooling where $\abs{\Omega_\mathrm{cool}} \gg |\Omega_{R/B}|$. The contribution of the on-resonance cooling sideband to the projective output signal filtered around $\Delta_\mathrm{proj.} \gg W$ can be included as a modification to the anti-Stokes coefficient,
\begin{equation}
    k_R \to k_R + \eta k_\mathrm{cool} e^{i\Delta_\mathrm{proj.} t},
\end{equation}
where
\begin{equation}
    \eta = \frac{1}{1-i\frac{2\Delta_\mathrm{proj.}}{W}},
\end{equation}
such that $|\eta|$ is the ratio of cooling sideband photons which are let through the filter. If this contribution is non-negligible such that the Stokes and anti-Stokes coefficients vary with time, one could instead formulate the postselection criterion in terms of an ideal projective measurement time $t_c$ such that
\begin{equation} \label{eq:large_cooling_contamination}
    \frac{k_R + \eta k_\mathrm{cool}e^{i\Delta_\mathrm{proj.} t_c}}{k_B} = \frac{r-\beta^*}{\beta}
\end{equation}
leads to the desired mechanical state.

Taking this principle to its extreme conclusion opens for an alternative approach: Rather than applying separate red-detuned drives $\Omega_R, \ \Omega_\mathrm{cool}$ for projective and readout measurements respectively, the cooling drive could serve as the anti-Stokes photon source for both measurements simultaneously. This would mean setting $k_R = 0$ in Equation \eqref{eq:large_cooling_contamination}. By tuning the drive magnitude $\abs{\Omega_B}$, one can then always find an ideal projective measurement time $t_c$ leading to the desired complex phase relationship, which is satisfied periodically with period $2\pi/\Delta_\mathrm{proj.}$.


\begin{thebibliography}{44}%
\makeatletter
\providecommand \@ifxundefined [1]{%
 \@ifx{#1\undefined}
}%
\providecommand \@ifnum [1]{%
 \ifnum #1\expandafter \@firstoftwo
 \else \expandafter \@secondoftwo
 \fi
}%
\providecommand \@ifx [1]{%
 \ifx #1\expandafter \@firstoftwo
 \else \expandafter \@secondoftwo
 \fi
}%
\providecommand \natexlab [1]{#1}%
\providecommand \enquote  [1]{``#1''}%
\providecommand \bibnamefont  [1]{#1}%
\providecommand \bibfnamefont [1]{#1}%
\providecommand \citenamefont [1]{#1}%
\providecommand \href@noop [0]{\@secondoftwo}%
\providecommand \href [0]{\begingroup \@sanitize@url \@href}%
\providecommand \@href[1]{\@@startlink{#1}\@@href}%
\providecommand \@@href[1]{\endgroup#1\@@endlink}%
\providecommand \@sanitize@url [0]{\catcode `\\12\catcode `\$12\catcode
  `\&12\catcode `\#12\catcode `\^12\catcode `\_12\catcode `\%12\relax}%
\providecommand \@@startlink[1]{}%
\providecommand \@@endlink[0]{}%
\providecommand \url  [0]{\begingroup\@sanitize@url \@url }%
\providecommand \@url [1]{\endgroup\@href {#1}{\urlprefix }}%
\providecommand \urlprefix  [0]{URL }%
\providecommand \Eprint [0]{\href }%
\providecommand \doibase [0]{https://doi.org/}%
\providecommand \selectlanguage [0]{\@gobble}%
\providecommand \bibinfo  [0]{\@secondoftwo}%
\providecommand \bibfield  [0]{\@secondoftwo}%
\providecommand \translation [1]{[#1]}%
\providecommand \BibitemOpen [0]{}%
\providecommand \bibitemStop [0]{}%
\providecommand \bibitemNoStop [0]{.\EOS\space}%
\providecommand \EOS [0]{\spacefactor3000\relax}%
\providecommand \BibitemShut  [1]{\csname bibitem#1\endcsname}%
\let\auto@bib@innerbib\@empty
\bibitem [{\citenamefont {Arndt}\ and\ \citenamefont
  {Hornberger}(2014)}]{Arndt2014NatPhys}%
  \BibitemOpen
  \bibfield  {author} {\bibinfo {author} {\bibfnamefont {M.}~\bibnamefont
  {Arndt}}\ and\ \bibinfo {author} {\bibfnamefont {K.}~\bibnamefont
  {Hornberger}},\ }\bibfield  {title} {\bibinfo {title} {Testing the limits of
  quantum mechanical superpositions},\ }\href@noop {} {\bibfield  {journal}
  {\bibinfo  {journal} {Nat. Phys}\ }\textbf {\bibinfo {volume} {10}},\
  \bibinfo {pages} {271} (\bibinfo {year} {2014})}\BibitemShut {NoStop}%
\bibitem [{\citenamefont {Bose}\ \emph {et~al.}(2017)\citenamefont {Bose},
  \citenamefont {Mazumdar}, \citenamefont {Morley}, \citenamefont {Ulbricht},
  \citenamefont {Toro\ifmmode~\check{s}\else \v{s}\fi{}}, \citenamefont
  {Paternostro}, \citenamefont {Geraci}, \citenamefont {Barker}, \citenamefont
  {Kim},\ and\ \citenamefont {Milburn}}]{Bose2017PRL}%
  \BibitemOpen
  \bibfield  {author} {\bibinfo {author} {\bibfnamefont {S.}~\bibnamefont
  {Bose}}, \bibinfo {author} {\bibfnamefont {A.}~\bibnamefont {Mazumdar}},
  \bibinfo {author} {\bibfnamefont {G.~W.}\ \bibnamefont {Morley}}, \bibinfo
  {author} {\bibfnamefont {H.}~\bibnamefont {Ulbricht}}, \bibinfo {author}
  {\bibfnamefont {M.}~\bibnamefont {Toro\ifmmode~\check{s}\else \v{s}\fi{}}},
  \bibinfo {author} {\bibfnamefont {M.}~\bibnamefont {Paternostro}}, \bibinfo
  {author} {\bibfnamefont {A.~A.}\ \bibnamefont {Geraci}}, \bibinfo {author}
  {\bibfnamefont {P.~F.}\ \bibnamefont {Barker}}, \bibinfo {author}
  {\bibfnamefont {M.~S.}\ \bibnamefont {Kim}},\ and\ \bibinfo {author}
  {\bibfnamefont {G.}~\bibnamefont {Milburn}},\ }\bibfield  {title} {\bibinfo
  {title} {Spin entanglement witness for quantum gravity},\ }\href
  {https://doi.org/10.1103/PhysRevLett.119.240401} {\bibfield  {journal}
  {\bibinfo  {journal} {Phys. Rev. Lett.}\ }\textbf {\bibinfo {volume} {119}},\
  \bibinfo {pages} {240401} (\bibinfo {year} {2017})}\BibitemShut {NoStop}%
\bibitem [{\citenamefont {Marletto}\ and\ \citenamefont
  {Vedral}(2017)}]{Marletto2017PRL}%
  \BibitemOpen
  \bibfield  {author} {\bibinfo {author} {\bibfnamefont {C.}~\bibnamefont
  {Marletto}}\ and\ \bibinfo {author} {\bibfnamefont {V.}~\bibnamefont
  {Vedral}},\ }\bibfield  {title} {\bibinfo {title} {Gravitationally induced
  entanglement between two massive particles is sufficient evidence of quantum
  effects in gravity},\ }\href {https://doi.org/10.1103/PhysRevLett.119.240402}
  {\bibfield  {journal} {\bibinfo  {journal} {Phys. Rev. Lett.}\ }\textbf
  {\bibinfo {volume} {119}},\ \bibinfo {pages} {240402} (\bibinfo {year}
  {2017})}\BibitemShut {NoStop}%
\bibitem [{\citenamefont {Leggett}(1980)}]{Leggett1980ProgTheorPhysSuppl}%
  \BibitemOpen
  \bibfield  {author} {\bibinfo {author} {\bibfnamefont {A.~J.}\ \bibnamefont
  {Leggett}},\ }\bibfield  {title} {\bibinfo {title} {{Macroscopic Quantum
  Systems and the Quantum Theory of Measurement}},\ }\href
  {https://doi.org/10.1143/PTP.69.80} {\bibfield  {journal} {\bibinfo
  {journal} {Progress of Theoretical Physics Supplement}\ }\textbf {\bibinfo
  {volume} {69}},\ \bibinfo {pages} {80} (\bibinfo {year} {1980})}\BibitemShut
  {NoStop}%
\bibitem [{\citenamefont {Fr\"owis}\ \emph {et~al.}(2018)\citenamefont
  {Fr\"owis}, \citenamefont {Sekatski}, \citenamefont {D\"ur}, \citenamefont
  {Gisin},\ and\ \citenamefont {Sangouard}}]{Frowis2018RMP}%
  \BibitemOpen
  \bibfield  {author} {\bibinfo {author} {\bibfnamefont {F.}~\bibnamefont
  {Fr\"owis}}, \bibinfo {author} {\bibfnamefont {P.}~\bibnamefont {Sekatski}},
  \bibinfo {author} {\bibfnamefont {W.}~\bibnamefont {D\"ur}}, \bibinfo
  {author} {\bibfnamefont {N.}~\bibnamefont {Gisin}},\ and\ \bibinfo {author}
  {\bibfnamefont {N.}~\bibnamefont {Sangouard}},\ }\bibfield  {title} {\bibinfo
  {title} {Macroscopic quantum states: Measures, fragility, and
  implementations},\ }\href {https://doi.org/10.1103/RevModPhys.90.025004}
  {\bibfield  {journal} {\bibinfo  {journal} {Rev. Mod. Phys.}\ }\textbf
  {\bibinfo {volume} {90}},\ \bibinfo {pages} {025004} (\bibinfo {year}
  {2018})}\BibitemShut {NoStop}%
\bibitem [{\citenamefont {Bruno}\ \emph {et~al.}(2013)\citenamefont {Bruno},
  \citenamefont {Martin}, \citenamefont {Sekatski}, \citenamefont {Sangouard},
  \citenamefont {Thew},\ and\ \citenamefont {Gisin}}]{Bruno2013NatPhys}%
  \BibitemOpen
  \bibfield  {author} {\bibinfo {author} {\bibfnamefont {N.}~\bibnamefont
  {Bruno}}, \bibinfo {author} {\bibfnamefont {A.}~\bibnamefont {Martin}},
  \bibinfo {author} {\bibfnamefont {P.}~\bibnamefont {Sekatski}}, \bibinfo
  {author} {\bibfnamefont {N.}~\bibnamefont {Sangouard}}, \bibinfo {author}
  {\bibfnamefont {R.~T.}\ \bibnamefont {Thew}},\ and\ \bibinfo {author}
  {\bibfnamefont {N.}~\bibnamefont {Gisin}},\ }\bibfield  {title} {\bibinfo
  {title} {{Displacement of entanglement back and forth between the micro and
  macro domains}},\ }\href {https://doi.org/10.1038/nphys2681} {\bibfield
  {journal} {\bibinfo  {journal} {Nature Physics}\ }\textbf {\bibinfo {volume}
  {9}},\ \bibinfo {pages} {545} (\bibinfo {year} {2013})}\BibitemShut {NoStop}%
\bibitem [{\citenamefont {Lvovsky}\ \emph {et~al.}(2013)\citenamefont
  {Lvovsky}, \citenamefont {Ghobadi}, \citenamefont {Chandra}, \citenamefont
  {Prasad},\ and\ \citenamefont {Simon}}]{Lvovsky2013NatPhys}%
  \BibitemOpen
  \bibfield  {author} {\bibinfo {author} {\bibfnamefont {A.~I.}\ \bibnamefont
  {Lvovsky}}, \bibinfo {author} {\bibfnamefont {R.}~\bibnamefont {Ghobadi}},
  \bibinfo {author} {\bibfnamefont {A.}~\bibnamefont {Chandra}}, \bibinfo
  {author} {\bibfnamefont {A.~S.}\ \bibnamefont {Prasad}},\ and\ \bibinfo
  {author} {\bibfnamefont {C.}~\bibnamefont {Simon}},\ }\bibfield  {title}
  {\bibinfo {title} {Observation of micro-macro entanglement of light},\
  }\href@noop {} {\bibfield  {journal} {\bibinfo  {journal} {Nat. Phys.}\
  }\textbf {\bibinfo {volume} {9}},\ \bibinfo {pages} {541} (\bibinfo {year}
  {2013})}\BibitemShut {NoStop}%
\bibitem [{\citenamefont {Sychev}\ \emph {et~al.}(2019)\citenamefont {Sychev},
  \citenamefont {Novikov}, \citenamefont {Pirov}, \citenamefont {Simon},\ and\
  \citenamefont {Lvovsky}}]{Sychev2019Optica}%
  \BibitemOpen
  \bibfield  {author} {\bibinfo {author} {\bibfnamefont {D.~V.}\ \bibnamefont
  {Sychev}}, \bibinfo {author} {\bibfnamefont {V.~A.}\ \bibnamefont {Novikov}},
  \bibinfo {author} {\bibfnamefont {K.~K.}\ \bibnamefont {Pirov}}, \bibinfo
  {author} {\bibfnamefont {C.}~\bibnamefont {Simon}},\ and\ \bibinfo {author}
  {\bibfnamefont {A.~I.}\ \bibnamefont {Lvovsky}},\ }\bibfield  {title}
  {\bibinfo {title} {Entanglement of macroscopically distinct states of
  light},\ }\href {https://doi.org/10.1364/OPTICA.6.001425} {\bibfield
  {journal} {\bibinfo  {journal} {Optica}\ }\textbf {\bibinfo {volume} {6}},\
  \bibinfo {pages} {1425} (\bibinfo {year} {2019})}\BibitemShut {NoStop}%
\bibitem [{\citenamefont {Biagi}\ \emph {et~al.}(2020)\citenamefont {Biagi},
  \citenamefont {Costanzo}, \citenamefont {Bellini},\ and\ \citenamefont
  {Zavatta}}]{Biagi2020PRL}%
  \BibitemOpen
  \bibfield  {author} {\bibinfo {author} {\bibfnamefont {N.}~\bibnamefont
  {Biagi}}, \bibinfo {author} {\bibfnamefont {L.~S.}\ \bibnamefont {Costanzo}},
  \bibinfo {author} {\bibfnamefont {M.}~\bibnamefont {Bellini}},\ and\ \bibinfo
  {author} {\bibfnamefont {A.}~\bibnamefont {Zavatta}},\ }\bibfield  {title}
  {\bibinfo {title} {Entangling macroscopic light states by delocalized photon
  addition},\ }\href {https://doi.org/10.1103/PhysRevLett.124.033604}
  {\bibfield  {journal} {\bibinfo  {journal} {Phys. Rev. Lett.}\ }\textbf
  {\bibinfo {volume} {124}},\ \bibinfo {pages} {033604} (\bibinfo {year}
  {2020})}\BibitemShut {NoStop}%
\bibitem [{\citenamefont {Sekatski}\ \emph {et~al.}(2014)\citenamefont
  {Sekatski}, \citenamefont {Sangouard},\ and\ \citenamefont
  {Gisin}}]{Sekatski2014PRA}%
  \BibitemOpen
  \bibfield  {author} {\bibinfo {author} {\bibfnamefont {P.}~\bibnamefont
  {Sekatski}}, \bibinfo {author} {\bibfnamefont {N.}~\bibnamefont
  {Sangouard}},\ and\ \bibinfo {author} {\bibfnamefont {N.}~\bibnamefont
  {Gisin}},\ }\bibfield  {title} {\bibinfo {title} {Size of quantum
  superpositions as measured with classical detectors},\ }\href@noop {}
  {\bibfield  {journal} {\bibinfo  {journal} {Physical Review A}\ }\textbf
  {\bibinfo {volume} {89}},\ \bibinfo {pages} {012116} (\bibinfo {year}
  {2014})}\BibitemShut {NoStop}%
\bibitem [{\citenamefont {de~Oliveira}\ \emph {et~al.}(1990)\citenamefont
  {de~Oliveira}, \citenamefont {Kim}, \citenamefont {Knight},\ and\
  \citenamefont {Buek}}]{Oliveira1990PRA}%
  \BibitemOpen
  \bibfield  {author} {\bibinfo {author} {\bibfnamefont {F.~A.~M.}\
  \bibnamefont {de~Oliveira}}, \bibinfo {author} {\bibfnamefont {M.~S.}\
  \bibnamefont {Kim}}, \bibinfo {author} {\bibfnamefont {P.~L.}\ \bibnamefont
  {Knight}},\ and\ \bibinfo {author} {\bibfnamefont {V.}~\bibnamefont {Buek}},\
  }\bibfield  {title} {\bibinfo {title} {{Properties of displaced number
  states}},\ }\href {https://doi.org/10.1103/PhysRevA.41.2645} {\bibfield
  {journal} {\bibinfo  {journal} {Phys. Rev. A}\ }\textbf {\bibinfo {volume}
  {41}},\ \bibinfo {pages} {2645} (\bibinfo {year} {1990})}\BibitemShut
  {NoStop}%
\bibitem [{\citenamefont {Aspelmeyer}\ \emph {et~al.}(2014)\citenamefont
  {Aspelmeyer}, \citenamefont {Kippenberg},\ and\ \citenamefont
  {Marquardt}}]{Aspelmeyer2014RMP}%
  \BibitemOpen
  \bibfield  {author} {\bibinfo {author} {\bibfnamefont {M.}~\bibnamefont
  {Aspelmeyer}}, \bibinfo {author} {\bibfnamefont {T.~J.}\ \bibnamefont
  {Kippenberg}},\ and\ \bibinfo {author} {\bibfnamefont {F.}~\bibnamefont
  {Marquardt}},\ }\bibfield  {title} {\bibinfo {title} {Cavity optomechanics},\
  }\href@noop {} {\bibfield  {journal} {\bibinfo  {journal} {Rev. Mod. Phys.}\
  }\textbf {\bibinfo {volume} {86}},\ \bibinfo {pages} {1391} (\bibinfo {year}
  {2014})}\BibitemShut {NoStop}%
\bibitem [{\citenamefont {Teufel}\ \emph {et~al.}(2011)\citenamefont {Teufel},
  \citenamefont {Donner}, \citenamefont {Li}, \citenamefont {Harlow},
  \citenamefont {Allman}, \citenamefont {Cicak}, \citenamefont {Sirois},
  \citenamefont {Whittaker}, \citenamefont {Lehnert},\ and\ \citenamefont
  {Simmonds}}]{Teufel2011Nature}%
  \BibitemOpen
  \bibfield  {author} {\bibinfo {author} {\bibfnamefont {J.~D.}\ \bibnamefont
  {Teufel}}, \bibinfo {author} {\bibfnamefont {T.}~\bibnamefont {Donner}},
  \bibinfo {author} {\bibfnamefont {D.}~\bibnamefont {Li}}, \bibinfo {author}
  {\bibfnamefont {J.~W.}\ \bibnamefont {Harlow}}, \bibinfo {author}
  {\bibfnamefont {M.~S.}\ \bibnamefont {Allman}}, \bibinfo {author}
  {\bibfnamefont {K.}~\bibnamefont {Cicak}}, \bibinfo {author} {\bibfnamefont
  {A.~J.}\ \bibnamefont {Sirois}}, \bibinfo {author} {\bibfnamefont {J.~D.}\
  \bibnamefont {Whittaker}}, \bibinfo {author} {\bibfnamefont {K.~W.}\
  \bibnamefont {Lehnert}},\ and\ \bibinfo {author} {\bibfnamefont {R.~W.}\
  \bibnamefont {Simmonds}},\ }\bibfield  {title} {\bibinfo {title} {Sideband
  cooling of micromechanical motion to the quantum ground state},\ }\href@noop
  {} {\bibfield  {journal} {\bibinfo  {journal} {Nature}\ }\textbf {\bibinfo
  {volume} {475}},\ \bibinfo {pages} {359} (\bibinfo {year}
  {2011})}\BibitemShut {NoStop}%
\bibitem [{\citenamefont {Chan}\ \emph {et~al.}(2011)\citenamefont {Chan},
  \citenamefont {Alegre}, \citenamefont {Safavi-Naeini}, \citenamefont {Hill},
  \citenamefont {Krause}, \citenamefont {Gr\"{o}blacher}, \citenamefont
  {Aspelmeyer},\ and\ \citenamefont {Painter}}]{Chan2011Nature}%
  \BibitemOpen
  \bibfield  {author} {\bibinfo {author} {\bibfnamefont {J.}~\bibnamefont
  {Chan}}, \bibinfo {author} {\bibfnamefont {T.~P.~M.}\ \bibnamefont {Alegre}},
  \bibinfo {author} {\bibfnamefont {A.~H.}\ \bibnamefont {Safavi-Naeini}},
  \bibinfo {author} {\bibfnamefont {J.~T.}\ \bibnamefont {Hill}}, \bibinfo
  {author} {\bibfnamefont {A.}~\bibnamefont {Krause}}, \bibinfo {author}
  {\bibfnamefont {S.}~\bibnamefont {Gr\"{o}blacher}}, \bibinfo {author}
  {\bibfnamefont {M.}~\bibnamefont {Aspelmeyer}},\ and\ \bibinfo {author}
  {\bibfnamefont {O.}~\bibnamefont {Painter}},\ }\bibfield  {title} {\bibinfo
  {title} {Laser cooling of a nanomechanical oscillator into its quantum ground
  state},\ }\href@noop {} {\bibfield  {journal} {\bibinfo  {journal} {Nature}\
  }\textbf {\bibinfo {volume} {478}},\ \bibinfo {pages} {89} (\bibinfo {year}
  {2011})}\BibitemShut {NoStop}%
\bibitem [{\citenamefont {Pirkkalainen}\ \emph {et~al.}(2015)\citenamefont
  {Pirkkalainen}, \citenamefont {Damsk\"agg}, \citenamefont {Brandt},
  \citenamefont {Massel},\ and\ \citenamefont
  {Sillanp\"a\"a}}]{Pirkkalainen2015PRL}%
  \BibitemOpen
  \bibfield  {author} {\bibinfo {author} {\bibfnamefont {J.-M.}\ \bibnamefont
  {Pirkkalainen}}, \bibinfo {author} {\bibfnamefont {E.}~\bibnamefont
  {Damsk\"agg}}, \bibinfo {author} {\bibfnamefont {M.}~\bibnamefont {Brandt}},
  \bibinfo {author} {\bibfnamefont {F.}~\bibnamefont {Massel}},\ and\ \bibinfo
  {author} {\bibfnamefont {M.~A.}\ \bibnamefont {Sillanp\"a\"a}},\ }\bibfield
  {title} {\bibinfo {title} {Squeezing of quantum noise of motion in a
  micromechanical resonator},\ }\href
  {https://doi.org/10.1103/PhysRevLett.115.243601} {\bibfield  {journal}
  {\bibinfo  {journal} {Phys. Rev. Lett.}\ }\textbf {\bibinfo {volume} {115}},\
  \bibinfo {pages} {243601} (\bibinfo {year} {2015})}\BibitemShut {NoStop}%
\bibitem [{\citenamefont {Lecocq}\ \emph {et~al.}(2015)\citenamefont {Lecocq},
  \citenamefont {Clark}, \citenamefont {Simmonds}, \citenamefont {Aumentado},\
  and\ \citenamefont {Teufel}}]{Lecocq2015PRX}%
  \BibitemOpen
  \bibfield  {author} {\bibinfo {author} {\bibfnamefont {F.}~\bibnamefont
  {Lecocq}}, \bibinfo {author} {\bibfnamefont {J.~B.}\ \bibnamefont {Clark}},
  \bibinfo {author} {\bibfnamefont {R.~W.}\ \bibnamefont {Simmonds}}, \bibinfo
  {author} {\bibfnamefont {J.}~\bibnamefont {Aumentado}},\ and\ \bibinfo
  {author} {\bibfnamefont {J.~D.}\ \bibnamefont {Teufel}},\ }\bibfield  {title}
  {\bibinfo {title} {Quantum nondemolition measurement of a nonclassical state
  of a massive object},\ }\href {https://doi.org/10.1103/PhysRevX.5.041037}
  {\bibfield  {journal} {\bibinfo  {journal} {Phys. Rev. X}\ }\textbf {\bibinfo
  {volume} {5}},\ \bibinfo {pages} {041037} (\bibinfo {year}
  {2015})}\BibitemShut {NoStop}%
\bibitem [{\citenamefont {Wollman}\ \emph {et~al.}(2015)\citenamefont
  {Wollman}, \citenamefont {Lei}, \citenamefont {Weinstein}, \citenamefont
  {Suh}, \citenamefont {Kronwald}, \citenamefont {Marquardt}, \citenamefont
  {Clerk},\ and\ \citenamefont {Schwab}}]{Wollman2015Science}%
  \BibitemOpen
  \bibfield  {author} {\bibinfo {author} {\bibfnamefont {E.~E.}\ \bibnamefont
  {Wollman}}, \bibinfo {author} {\bibfnamefont {C.~U.}\ \bibnamefont {Lei}},
  \bibinfo {author} {\bibfnamefont {A.~J.}\ \bibnamefont {Weinstein}}, \bibinfo
  {author} {\bibfnamefont {J.}~\bibnamefont {Suh}}, \bibinfo {author}
  {\bibfnamefont {A.}~\bibnamefont {Kronwald}}, \bibinfo {author}
  {\bibfnamefont {F.}~\bibnamefont {Marquardt}}, \bibinfo {author}
  {\bibfnamefont {A.~A.}\ \bibnamefont {Clerk}},\ and\ \bibinfo {author}
  {\bibfnamefont {K.~C.}\ \bibnamefont {Schwab}},\ }\bibfield  {title}
  {\bibinfo {title} {Quantum squeezing of motion in a mechanical resonator},\
  }\href {https://doi.org/10.1126/science.aac5138} {\bibfield  {journal}
  {\bibinfo  {journal} {Science}\ }\textbf {\bibinfo {volume} {349}},\ \bibinfo
  {pages} {952} (\bibinfo {year} {2015})}\BibitemShut {NoStop}%
\bibitem [{\citenamefont {Wang}(2023)}]{Wang2023PhDThesis}%
  \BibitemOpen
  \bibfield  {author} {\bibinfo {author} {\bibfnamefont {Y.}~\bibnamefont
  {Wang}},\ }\emph {\bibinfo {title} {Manipulating and Measuring States of a
  Superfluid Optomechanical Resonator in the Quantum Regime}},\ \href
  {https://harrislab.yale.edu/files/thesis/Wang_Thesis.pdf} {Ph.D. thesis},\
  \bibinfo  {school} {Yale University} (\bibinfo {year} {2023})\BibitemShut
  {NoStop}%
\bibitem [{\citenamefont {Riedinger}\ \emph {et~al.}(2016)\citenamefont
  {Riedinger}, \citenamefont {Hong}, \citenamefont {Norte}, \citenamefont
  {Slater}, \citenamefont {Shang}, \citenamefont {Krause}, \citenamefont
  {Anant}, \citenamefont {Aspelmeyer},\ and\ \citenamefont
  {Gr\"{o}blacher}}]{Riedinger2016Nature}%
  \BibitemOpen
  \bibfield  {author} {\bibinfo {author} {\bibfnamefont {R.}~\bibnamefont
  {Riedinger}}, \bibinfo {author} {\bibfnamefont {S.}~\bibnamefont {Hong}},
  \bibinfo {author} {\bibfnamefont {R.~A.}\ \bibnamefont {Norte}}, \bibinfo
  {author} {\bibfnamefont {J.~A.}\ \bibnamefont {Slater}}, \bibinfo {author}
  {\bibfnamefont {J.}~\bibnamefont {Shang}}, \bibinfo {author} {\bibfnamefont
  {A.~G.}\ \bibnamefont {Krause}}, \bibinfo {author} {\bibfnamefont
  {V.}~\bibnamefont {Anant}}, \bibinfo {author} {\bibfnamefont
  {M.}~\bibnamefont {Aspelmeyer}},\ and\ \bibinfo {author} {\bibfnamefont
  {S.}~\bibnamefont {Gr\"{o}blacher}},\ }\bibfield  {title} {\bibinfo {title}
  {Non-classical correlations between single photons and phonons from a
  mechanical oscillator},\ }\href@noop {} {\bibfield  {journal} {\bibinfo
  {journal} {Nature}\ }\textbf {\bibinfo {volume} {530}},\ \bibinfo {pages}
  {313} (\bibinfo {year} {2016})}\BibitemShut {NoStop}%
\bibitem [{\citenamefont {Hong}\ \emph {et~al.}(2017)\citenamefont {Hong},
  \citenamefont {Riedinger}, \citenamefont {Marinkovi{\'c}}, \citenamefont
  {Wallucks}, \citenamefont {Hofer}, \citenamefont {Norte}, \citenamefont
  {Aspelmeyer},\ and\ \citenamefont {Gr{\"o}blacher}}]{Hong2017Science}%
  \BibitemOpen
  \bibfield  {author} {\bibinfo {author} {\bibfnamefont {S.}~\bibnamefont
  {Hong}}, \bibinfo {author} {\bibfnamefont {R.}~\bibnamefont {Riedinger}},
  \bibinfo {author} {\bibfnamefont {I.}~\bibnamefont {Marinkovi{\'c}}},
  \bibinfo {author} {\bibfnamefont {A.}~\bibnamefont {Wallucks}}, \bibinfo
  {author} {\bibfnamefont {S.~G.}\ \bibnamefont {Hofer}}, \bibinfo {author}
  {\bibfnamefont {R.~A.}\ \bibnamefont {Norte}}, \bibinfo {author}
  {\bibfnamefont {M.}~\bibnamefont {Aspelmeyer}},\ and\ \bibinfo {author}
  {\bibfnamefont {S.}~\bibnamefont {Gr{\"o}blacher}},\ }\bibfield  {title}
  {\bibinfo {title} {Hanbury-{B}rown and {T}wiss interferometry of single
  phonons from an optomechanical resonator},\ }\href
  {https://doi.org/10.1126/science.aan7939} {\bibfield  {journal} {\bibinfo
  {journal} {Science}\ }\textbf {\bibinfo {volume} {358}},\ \bibinfo {pages}
  {203} (\bibinfo {year} {2017})}\BibitemShut {NoStop}%
\bibitem [{\citenamefont {Velez}\ \emph {et~al.}(2019)\citenamefont {Velez},
  \citenamefont {Seibold}, \citenamefont {Kipfer}, \citenamefont {Anderson},
  \citenamefont {Sudhir},\ and\ \citenamefont {Galland}}]{Velez2019PRX}%
  \BibitemOpen
  \bibfield  {author} {\bibinfo {author} {\bibfnamefont {S.~T.}\ \bibnamefont
  {Velez}}, \bibinfo {author} {\bibfnamefont {K.}~\bibnamefont {Seibold}},
  \bibinfo {author} {\bibfnamefont {N.}~\bibnamefont {Kipfer}}, \bibinfo
  {author} {\bibfnamefont {M.~D.}\ \bibnamefont {Anderson}}, \bibinfo {author}
  {\bibfnamefont {V.}~\bibnamefont {Sudhir}},\ and\ \bibinfo {author}
  {\bibfnamefont {C.}~\bibnamefont {Galland}},\ }\bibfield  {title} {\bibinfo
  {title} {Preparation and decay of a single quantum of vibration at ambient
  conditions},\ }\href {https://doi.org/10.1103/PhysRevX.9.041007} {\bibfield
  {journal} {\bibinfo  {journal} {Phys. Rev. X}\ }\textbf {\bibinfo {volume}
  {9}},\ \bibinfo {pages} {041007} (\bibinfo {year} {2019})}\BibitemShut
  {NoStop}%
\bibitem [{\citenamefont {Lee}\ \emph {et~al.}(2011)\citenamefont {Lee},
  \citenamefont {Sprague}, \citenamefont {Sussman}, \citenamefont {Nunn},
  \citenamefont {Langford}, \citenamefont {Jin}, \citenamefont {Champion},
  \citenamefont {Michelberger}, \citenamefont {Reim}, \citenamefont {England},
  \citenamefont {Jaksch},\ and\ \citenamefont {Walmsley}}]{Lee2011Science}%
  \BibitemOpen
  \bibfield  {author} {\bibinfo {author} {\bibfnamefont {K.~C.}\ \bibnamefont
  {Lee}}, \bibinfo {author} {\bibfnamefont {M.~R.}\ \bibnamefont {Sprague}},
  \bibinfo {author} {\bibfnamefont {B.~J.}\ \bibnamefont {Sussman}}, \bibinfo
  {author} {\bibfnamefont {J.}~\bibnamefont {Nunn}}, \bibinfo {author}
  {\bibfnamefont {N.~K.}\ \bibnamefont {Langford}}, \bibinfo {author}
  {\bibfnamefont {X.-M.}\ \bibnamefont {Jin}}, \bibinfo {author} {\bibfnamefont
  {T.}~\bibnamefont {Champion}}, \bibinfo {author} {\bibfnamefont
  {P.}~\bibnamefont {Michelberger}}, \bibinfo {author} {\bibfnamefont {K.~F.}\
  \bibnamefont {Reim}}, \bibinfo {author} {\bibfnamefont {D.}~\bibnamefont
  {England}}, \bibinfo {author} {\bibfnamefont {D.}~\bibnamefont {Jaksch}},\
  and\ \bibinfo {author} {\bibfnamefont {I.~A.}\ \bibnamefont {Walmsley}},\
  }\bibfield  {title} {\bibinfo {title} {Entangling macroscopic diamonds at
  room temperature},\ }\href {https://doi.org/10.1126/science.1211914}
  {\bibfield  {journal} {\bibinfo  {journal} {Science}\ }\textbf {\bibinfo
  {volume} {334}},\ \bibinfo {pages} {1253} (\bibinfo {year}
  {2011})}\BibitemShut {NoStop}%
\bibitem [{\citenamefont {Cohen}\ \emph {et~al.}(2015)\citenamefont {Cohen},
  \citenamefont {Meenehan}, \citenamefont {MacCabe}, \citenamefont
  {Gr\"{o}blacher}, \citenamefont {Safavi-Naeini}, \citenamefont {Marsili},
  \citenamefont {Shaw},\ and\ \citenamefont {Painter}}]{Cohen2015Nature}%
  \BibitemOpen
  \bibfield  {author} {\bibinfo {author} {\bibfnamefont {J.~D.}\ \bibnamefont
  {Cohen}}, \bibinfo {author} {\bibfnamefont {S.~M.}\ \bibnamefont {Meenehan}},
  \bibinfo {author} {\bibfnamefont {G.~S.}\ \bibnamefont {MacCabe}}, \bibinfo
  {author} {\bibfnamefont {S.}~\bibnamefont {Gr\"{o}blacher}}, \bibinfo
  {author} {\bibfnamefont {A.~H.}\ \bibnamefont {Safavi-Naeini}}, \bibinfo
  {author} {\bibfnamefont {F.}~\bibnamefont {Marsili}}, \bibinfo {author}
  {\bibfnamefont {M.~D.}\ \bibnamefont {Shaw}},\ and\ \bibinfo {author}
  {\bibfnamefont {O.}~\bibnamefont {Painter}},\ }\bibfield  {title} {\bibinfo
  {title} {Phonon counting and intensity interferometry of a nanomechanical
  resonator},\ }\href@noop {} {\bibfield  {journal} {\bibinfo  {journal}
  {Nature}\ }\textbf {\bibinfo {volume} {520}},\ \bibinfo {pages} {522}
  (\bibinfo {year} {2015})}\BibitemShut {NoStop}%
\bibitem [{\citenamefont {Patel}\ \emph {et~al.}(2021)\citenamefont {Patel},
  \citenamefont {McKenna}, \citenamefont {Wang}, \citenamefont {Witmer},
  \citenamefont {Jiang}, \citenamefont {Van~Laer}, \citenamefont {Sarabalis},\
  and\ \citenamefont {Safavi-Naeini}}]{Patel2021PRL}%
  \BibitemOpen
  \bibfield  {author} {\bibinfo {author} {\bibfnamefont {R.~N.}\ \bibnamefont
  {Patel}}, \bibinfo {author} {\bibfnamefont {T.~P.}\ \bibnamefont {McKenna}},
  \bibinfo {author} {\bibfnamefont {Z.}~\bibnamefont {Wang}}, \bibinfo {author}
  {\bibfnamefont {J.~D.}\ \bibnamefont {Witmer}}, \bibinfo {author}
  {\bibfnamefont {W.}~\bibnamefont {Jiang}}, \bibinfo {author} {\bibfnamefont
  {R.}~\bibnamefont {Van~Laer}}, \bibinfo {author} {\bibfnamefont {C.~J.}\
  \bibnamefont {Sarabalis}},\ and\ \bibinfo {author} {\bibfnamefont {A.~H.}\
  \bibnamefont {Safavi-Naeini}},\ }\bibfield  {title} {\bibinfo {title}
  {Room-temperature mechanical resonator with a single added or subtracted
  phonon},\ }\href {https://doi.org/10.1103/PhysRevLett.127.133602} {\bibfield
  {journal} {\bibinfo  {journal} {Phys. Rev. Lett.}\ }\textbf {\bibinfo
  {volume} {127}},\ \bibinfo {pages} {133602} (\bibinfo {year}
  {2021})}\BibitemShut {NoStop}%
\bibitem [{\citenamefont {Galinskiy}\ \emph {et~al.}(2020)\citenamefont
  {Galinskiy}, \citenamefont {Tsaturyan}, \citenamefont {Parniak},\ and\
  \citenamefont {Polzik}}]{Galinskiy2020Optica}%
  \BibitemOpen
  \bibfield  {author} {\bibinfo {author} {\bibfnamefont {I.}~\bibnamefont
  {Galinskiy}}, \bibinfo {author} {\bibfnamefont {Y.}~\bibnamefont
  {Tsaturyan}}, \bibinfo {author} {\bibfnamefont {M.}~\bibnamefont {Parniak}},\
  and\ \bibinfo {author} {\bibfnamefont {E.~S.}\ \bibnamefont {Polzik}},\
  }\bibfield  {title} {\bibinfo {title} {Phonon counting thermometry of an
  ultracoherent membrane resonator near its motional ground state},\ }\href
  {https://doi.org/10.1364/OPTICA.390939} {\bibfield  {journal} {\bibinfo
  {journal} {Optica}\ }\textbf {\bibinfo {volume} {7}},\ \bibinfo {pages} {718}
  (\bibinfo {year} {2020})}\BibitemShut {NoStop}%
\bibitem [{\citenamefont {Enzian}\ \emph
  {et~al.}(2021{\natexlab{a}})\citenamefont {Enzian}, \citenamefont {Price},
  \citenamefont {Freisem}, \citenamefont {Nunn}, \citenamefont {Janousek},
  \citenamefont {Buchler}, \citenamefont {Lam},\ and\ \citenamefont
  {Vanner}}]{Enzian2021PRL}%
  \BibitemOpen
  \bibfield  {author} {\bibinfo {author} {\bibfnamefont {G.}~\bibnamefont
  {Enzian}}, \bibinfo {author} {\bibfnamefont {J.~J.}\ \bibnamefont {Price}},
  \bibinfo {author} {\bibfnamefont {L.}~\bibnamefont {Freisem}}, \bibinfo
  {author} {\bibfnamefont {J.}~\bibnamefont {Nunn}}, \bibinfo {author}
  {\bibfnamefont {J.}~\bibnamefont {Janousek}}, \bibinfo {author}
  {\bibfnamefont {B.~C.}\ \bibnamefont {Buchler}}, \bibinfo {author}
  {\bibfnamefont {P.~K.}\ \bibnamefont {Lam}},\ and\ \bibinfo {author}
  {\bibfnamefont {M.~R.}\ \bibnamefont {Vanner}},\ }\bibfield  {title}
  {\bibinfo {title} {Single-phonon addition and subtraction to a mechanical
  thermal state},\ }\href {https://doi.org/10.1103/PhysRevLett.126.033601}
  {\bibfield  {journal} {\bibinfo  {journal} {Phys. Rev. Lett.}\ }\textbf
  {\bibinfo {volume} {126}},\ \bibinfo {pages} {033601} (\bibinfo {year}
  {2021}{\natexlab{a}})}\BibitemShut {NoStop}%
\bibitem [{\citenamefont {Enzian}\ \emph
  {et~al.}(2021{\natexlab{b}})\citenamefont {Enzian}, \citenamefont {Freisem},
  \citenamefont {Price}, \citenamefont {Svela}, \citenamefont {Clarke},
  \citenamefont {Shajilal}, \citenamefont {Janousek}, \citenamefont {Buchler},
  \citenamefont {Lam},\ and\ \citenamefont {Vanner}}]{Enzian2021PRL_2}%
  \BibitemOpen
  \bibfield  {author} {\bibinfo {author} {\bibfnamefont {G.}~\bibnamefont
  {Enzian}}, \bibinfo {author} {\bibfnamefont {L.}~\bibnamefont {Freisem}},
  \bibinfo {author} {\bibfnamefont {J.~J.}\ \bibnamefont {Price}}, \bibinfo
  {author} {\bibfnamefont {A.~O.}\ \bibnamefont {Svela}}, \bibinfo {author}
  {\bibfnamefont {J.}~\bibnamefont {Clarke}}, \bibinfo {author} {\bibfnamefont
  {B.}~\bibnamefont {Shajilal}}, \bibinfo {author} {\bibfnamefont
  {J.}~\bibnamefont {Janousek}}, \bibinfo {author} {\bibfnamefont {B.~C.}\
  \bibnamefont {Buchler}}, \bibinfo {author} {\bibfnamefont {P.~K.}\
  \bibnamefont {Lam}},\ and\ \bibinfo {author} {\bibfnamefont {M.~R.}\
  \bibnamefont {Vanner}},\ }\bibfield  {title} {\bibinfo {title} {Non-gaussian
  mechanical motion via single and multiphonon subtraction from a thermal
  state},\ }\href {https://doi.org/10.1103/PhysRevLett.127.243601} {\bibfield
  {journal} {\bibinfo  {journal} {Phys. Rev. Lett.}\ }\textbf {\bibinfo
  {volume} {127}},\ \bibinfo {pages} {243601} (\bibinfo {year}
  {2021}{\natexlab{b}})}\BibitemShut {NoStop}%
\bibitem [{\citenamefont {Patil}\ \emph {et~al.}(2022)\citenamefont {Patil},
  \citenamefont {Yu}, \citenamefont {Frazier}, \citenamefont {Wang},
  \citenamefont {Johnson}, \citenamefont {Fox}, \citenamefont {Reichel},\ and\
  \citenamefont {Harris}}]{Patil2022PRL}%
  \BibitemOpen
  \bibfield  {author} {\bibinfo {author} {\bibfnamefont {Y.~S.~S.}\
  \bibnamefont {Patil}}, \bibinfo {author} {\bibfnamefont {J.}~\bibnamefont
  {Yu}}, \bibinfo {author} {\bibfnamefont {S.}~\bibnamefont {Frazier}},
  \bibinfo {author} {\bibfnamefont {Y.}~\bibnamefont {Wang}}, \bibinfo {author}
  {\bibfnamefont {K.}~\bibnamefont {Johnson}}, \bibinfo {author} {\bibfnamefont
  {J.}~\bibnamefont {Fox}}, \bibinfo {author} {\bibfnamefont {J.}~\bibnamefont
  {Reichel}},\ and\ \bibinfo {author} {\bibfnamefont {J.~G.~E.}\ \bibnamefont
  {Harris}},\ }\bibfield  {title} {\bibinfo {title} {Measuring high-order
  phonon correlations in an optomechanical resonator},\ }\href
  {https://doi.org/10.1103/PhysRevLett.128.183601} {\bibfield  {journal}
  {\bibinfo  {journal} {Phys. Rev. Lett.}\ }\textbf {\bibinfo {volume} {128}},\
  \bibinfo {pages} {183601} (\bibinfo {year} {2022})}\BibitemShut {NoStop}%
\bibitem [{\citenamefont {Milburn}\ \emph {et~al.}(2016)\citenamefont
  {Milburn}, \citenamefont {Kim},\ and\ \citenamefont
  {Vanner}}]{Milburn2016PRA}%
  \BibitemOpen
  \bibfield  {author} {\bibinfo {author} {\bibfnamefont {T.~J.}\ \bibnamefont
  {Milburn}}, \bibinfo {author} {\bibfnamefont {M.~S.}\ \bibnamefont {Kim}},\
  and\ \bibinfo {author} {\bibfnamefont {M.~R.}\ \bibnamefont {Vanner}},\
  }\bibfield  {title} {\bibinfo {title} {{Nonclassical-state generation in
  macroscopic systems via hybrid discrete-continuous quantum measurements}},\
  }\href {https://doi.org/10.1103/PhysRevA.93.053818} {\bibfield  {journal}
  {\bibinfo  {journal} {Phys. Rev. A}\ }\textbf {\bibinfo {volume} {93}},\
  \bibinfo {pages} {53818} (\bibinfo {year} {2016})}\BibitemShut {NoStop}%
\bibitem [{\citenamefont {Shomroni}\ \emph {et~al.}(2020)\citenamefont
  {Shomroni}, \citenamefont {Qiu},\ and\ \citenamefont
  {Kippenberg}}]{Shomroni2020PRA}%
  \BibitemOpen
  \bibfield  {author} {\bibinfo {author} {\bibfnamefont {I.}~\bibnamefont
  {Shomroni}}, \bibinfo {author} {\bibfnamefont {L.}~\bibnamefont {Qiu}},\ and\
  \bibinfo {author} {\bibfnamefont {T.~J.}\ \bibnamefont {Kippenberg}},\
  }\bibfield  {title} {\bibinfo {title} {{Optomechanical generation of a
  mechanical catlike state by phonon subtraction}},\ }\href
  {https://doi.org/10.1103/PhysRevA.101.033812} {\bibfield  {journal} {\bibinfo
   {journal} {Phys. Rev. A}\ }\textbf {\bibinfo {volume} {101}},\ \bibinfo
  {pages} {33812} (\bibinfo {year} {2020})}\BibitemShut {NoStop}%
\bibitem [{\citenamefont {Zhan}\ \emph {et~al.}(2020)\citenamefont {Zhan},
  \citenamefont {Li},\ and\ \citenamefont {Tan}}]{Zhan2020PRA}%
  \BibitemOpen
  \bibfield  {author} {\bibinfo {author} {\bibfnamefont {H.}~\bibnamefont
  {Zhan}}, \bibinfo {author} {\bibfnamefont {G.}~\bibnamefont {Li}},\ and\
  \bibinfo {author} {\bibfnamefont {H.}~\bibnamefont {Tan}},\ }\bibfield
  {title} {\bibinfo {title} {Preparing macroscopic mechanical quantum
  superpositions via photon detection},\ }\href
  {https://doi.org/10.1103/PhysRevA.101.063834} {\bibfield  {journal} {\bibinfo
   {journal} {Phys. Rev. A}\ }\textbf {\bibinfo {volume} {101}},\ \bibinfo
  {pages} {063834} (\bibinfo {year} {2020})}\BibitemShut {NoStop}%
\bibitem [{\citenamefont {Li}\ \emph {et~al.}(2018)\citenamefont {Li},
  \citenamefont {Gr{\"{o}}blacher}, \citenamefont {Zhu},\ and\ \citenamefont
  {Agarwal}}]{Li2018PRA}%
  \BibitemOpen
  \bibfield  {author} {\bibinfo {author} {\bibfnamefont {J.}~\bibnamefont
  {Li}}, \bibinfo {author} {\bibfnamefont {S.}~\bibnamefont
  {Gr{\"{o}}blacher}}, \bibinfo {author} {\bibfnamefont {S.-Y.}\ \bibnamefont
  {Zhu}},\ and\ \bibinfo {author} {\bibfnamefont {G.~S.}\ \bibnamefont
  {Agarwal}},\ }\bibfield  {title} {\bibinfo {title} {{Generation and detection
  of non-Gaussian phonon-added coherent states in optomechanical systems}},\
  }\href {https://doi.org/10.1103/PhysRevA.98.011801} {\bibfield  {journal}
  {\bibinfo  {journal} {Phys. Rev. A}\ }\textbf {\bibinfo {volume} {98}},\
  \bibinfo {pages} {11801} (\bibinfo {year} {2018})}\BibitemShut {NoStop}%
\bibitem [{\citenamefont {Vanner}\ \emph {et~al.}(2013)\citenamefont {Vanner},
  \citenamefont {Aspelmeyer},\ and\ \citenamefont {Kim}}]{Vanner2013PRL}%
  \BibitemOpen
  \bibfield  {author} {\bibinfo {author} {\bibfnamefont {M.~R.}\ \bibnamefont
  {Vanner}}, \bibinfo {author} {\bibfnamefont {M.}~\bibnamefont {Aspelmeyer}},\
  and\ \bibinfo {author} {\bibfnamefont {M.~S.}\ \bibnamefont {Kim}},\
  }\bibfield  {title} {\bibinfo {title} {Quantum state orthogonalization and a
  toolset for quantum optomechanical phonon control},\ }\href
  {https://doi.org/10.1103/PhysRevLett.110.010504} {\bibfield  {journal}
  {\bibinfo  {journal} {Phys. Rev. Lett.}\ }\textbf {\bibinfo {volume} {110}},\
  \bibinfo {pages} {010504} (\bibinfo {year} {2013})}\BibitemShut {NoStop}%
\bibitem [{\citenamefont {B{\o}rkje}\ \emph {et~al.}(2021)\citenamefont
  {B{\o}rkje}, \citenamefont {Massel},\ and\ \citenamefont
  {Harris}}]{borkje2021nonclassical}%
  \BibitemOpen
  \bibfield  {author} {\bibinfo {author} {\bibfnamefont {K.}~\bibnamefont
  {B{\o}rkje}}, \bibinfo {author} {\bibfnamefont {F.}~\bibnamefont {Massel}},\
  and\ \bibinfo {author} {\bibfnamefont {J.}~\bibnamefont {Harris}},\
  }\bibfield  {title} {\bibinfo {title} {Nonclassical photon statistics in
  two-tone continuously driven optomechanics},\ }\href@noop {} {\bibfield
  {journal} {\bibinfo  {journal} {Physical Review A}\ }\textbf {\bibinfo
  {volume} {104}},\ \bibinfo {pages} {063507} (\bibinfo {year}
  {2021})}\BibitemShut {NoStop}%
 \bibitem [{\citenamefont {Moya-Cessa}(1995)}]{Moya-Cessa1995JModOpt}%
  \BibitemOpen
  \bibfield  {author} {\bibinfo {author} {\bibfnamefont {H.}\ \bibnamefont
  {Moya-Cessa}},\ }\bibfield  {title} {\bibinfo {title} {{Generation and properties of superpositions of displaced Fock states}},\ } {\bibfield  {journal} {\bibinfo
  {journal} {Journal of Modern Optics}\ }\textbf {\bibinfo
  {volume} {42}},\ \bibinfo {pages} {1741} (\bibinfo {year} {1995})}\BibitemShut
  {NoStop}%
\bibitem [{\citenamefont {Lee}\ and\ \citenamefont
  {Nha}(2010)}]{lee2010quantum}%
  \BibitemOpen
  \bibfield  {author} {\bibinfo {author} {\bibfnamefont {S.-Y.}\ \bibnamefont
  {Lee}}\ and\ \bibinfo {author} {\bibfnamefont {H.}~\bibnamefont {Nha}},\
  }\bibfield  {title} {\bibinfo {title} {Quantum state engineering by a
  coherent superposition of photon subtraction and addition},\ }\href@noop {}
  {\bibfield  {journal} {\bibinfo  {journal} {Physical Review A}\ }\textbf
  {\bibinfo {volume} {82}},\ \bibinfo {pages} {053812} (\bibinfo {year}
  {2010})}\BibitemShut {NoStop}%
\bibitem [{\citenamefont {Rahim}\ \emph {et~al.}(2022)\citenamefont {Rahim},
  \citenamefont {Ooi},\ and\ \citenamefont {Othman}}]{rahim2022enhancing}%
  \BibitemOpen
  \bibfield  {author} {\bibinfo {author} {\bibfnamefont {M.}~\bibnamefont
  {Rahim}}, \bibinfo {author} {\bibfnamefont {C.~R.}\ \bibnamefont {Ooi}},\
  and\ \bibinfo {author} {\bibfnamefont {M.}~\bibnamefont {Othman}},\
  }\bibfield  {title} {\bibinfo {title} {Enhancing non-classicality by
  superposing two induced states from coherent states},\ }\href@noop {}
  {\bibfield  {journal} {\bibinfo  {journal} {International Journal of
  Theoretical Physics}\ }\textbf {\bibinfo {volume} {61}},\ \bibinfo {pages}
  {264} (\bibinfo {year} {2022})}\BibitemShut {NoStop}%
\bibitem [{\citenamefont {Viola~Kusminskiy}\ \emph {et~al.}(2016)\citenamefont
  {Viola~Kusminskiy}, \citenamefont {Tang},\ and\ \citenamefont
  {Marquardt}}]{Kusminskiy2016PRA}%
  \BibitemOpen
  \bibfield  {author} {\bibinfo {author} {\bibfnamefont {S.}~\bibnamefont
  {Viola~Kusminskiy}}, \bibinfo {author} {\bibfnamefont {H.~X.}\ \bibnamefont
  {Tang}},\ and\ \bibinfo {author} {\bibfnamefont {F.}~\bibnamefont
  {Marquardt}},\ }\bibfield  {title} {\bibinfo {title} {Coupled spin-light
  dynamics in cavity optomagnonics},\ }\href
  {https://doi.org/10.1103/PhysRevA.94.033821} {\bibfield  {journal} {\bibinfo
  {journal} {Phys. Rev. A}\ }\textbf {\bibinfo {volume} {94}},\ \bibinfo
  {pages} {033821} (\bibinfo {year} {2016})}\BibitemShut {NoStop}%
\bibitem [{\citenamefont {Bittencourt}\ \emph {et~al.}(2019)\citenamefont
  {Bittencourt}, \citenamefont {Feulner},\ and\ \citenamefont
  {Kusminskiy}}]{Bittencourt2019PRA}%
  \BibitemOpen
  \bibfield  {author} {\bibinfo {author} {\bibfnamefont {V.~A. S.~V.}\
  \bibnamefont {Bittencourt}}, \bibinfo {author} {\bibfnamefont
  {V.}~\bibnamefont {Feulner}},\ and\ \bibinfo {author} {\bibfnamefont {S.~V.}\
  \bibnamefont {Kusminskiy}},\ }\bibfield  {title} {\bibinfo {title} {Magnon
  heralding in cavity optomagnonics},\ }\href
  {https://doi.org/10.1103/PhysRevA.100.013810} {\bibfield  {journal} {\bibinfo
   {journal} {Phys. Rev. A}\ }\textbf {\bibinfo {volume} {100}},\ \bibinfo
  {pages} {013810} (\bibinfo {year} {2019})}\BibitemShut {NoStop}%
\bibitem [{\citenamefont {Mandel}(1979)}]{Mandel:79}%
  \BibitemOpen
  \bibfield  {author} {\bibinfo {author} {\bibfnamefont {L.}~\bibnamefont
  {Mandel}},\ }\bibfield  {title} {\bibinfo {title} {Sub-poissonian photon
  statistics in resonance fluorescence},\ }\href
  {https://doi.org/10.1364/OL.4.000205} {\bibfield  {journal} {\bibinfo
  {journal} {Opt. Lett.}\ }\textbf {\bibinfo {volume} {4}},\ \bibinfo {pages}
  {205} (\bibinfo {year} {1979})}\BibitemShut {NoStop}%
\bibitem [{Sup()}]{Supplementary}%
  \BibitemOpen
  \href@noop {} {}\bibinfo {note} {See Supplementary Material}\BibitemShut
  {NoStop}%
\bibitem [{\citenamefont {Bonaldi}\ \emph {et~al.}(2020)\citenamefont
  {Bonaldi}, \citenamefont {Borrielli}, \citenamefont {Chowdhury},
  \citenamefont {Di~Giuseppe}, \citenamefont {Li}, \citenamefont {Malossi},
  \citenamefont {Marino}, \citenamefont {Morana}, \citenamefont {Natali},
  \citenamefont {Piergentili}, \citenamefont {Prodi}, \citenamefont {Sarro},
  \citenamefont {Serra}, \citenamefont {Vezio}, \citenamefont {Vitali},\ and\
  \citenamefont {Marin}}]{Bonaldi2020EurPhysJD}%
  \BibitemOpen
  \bibfield  {author} {\bibinfo {author} {\bibfnamefont {M.}~\bibnamefont
  {Bonaldi}}, \bibinfo {author} {\bibfnamefont {A.}~\bibnamefont {Borrielli}},
  \bibinfo {author} {\bibfnamefont {A.}~\bibnamefont {Chowdhury}}, \bibinfo
  {author} {\bibfnamefont {G.}~\bibnamefont {Di~Giuseppe}}, \bibinfo {author}
  {\bibfnamefont {W.}~\bibnamefont {Li}}, \bibinfo {author} {\bibfnamefont
  {N.}~\bibnamefont {Malossi}}, \bibinfo {author} {\bibfnamefont
  {F.}~\bibnamefont {Marino}}, \bibinfo {author} {\bibfnamefont
  {B.}~\bibnamefont {Morana}}, \bibinfo {author} {\bibfnamefont
  {R.}~\bibnamefont {Natali}}, \bibinfo {author} {\bibfnamefont
  {P.}~\bibnamefont {Piergentili}}, \bibinfo {author} {\bibfnamefont {G.~A.}\
  \bibnamefont {Prodi}}, \bibinfo {author} {\bibfnamefont {P.~M.}\ \bibnamefont
  {Sarro}}, \bibinfo {author} {\bibfnamefont {E.}~\bibnamefont {Serra}},
  \bibinfo {author} {\bibfnamefont {P.}~\bibnamefont {Vezio}}, \bibinfo
  {author} {\bibfnamefont {D.}~\bibnamefont {Vitali}},\ and\ \bibinfo {author}
  {\bibfnamefont {F.}~\bibnamefont {Marin}},\ }\bibfield  {title} {\bibinfo
  {title} {Probing quantum gravity effects with quantum mechanical
  oscillators},\ }\href@noop {} {\bibfield  {journal} {\bibinfo  {journal}
  {Eur. Phys. J. D}\ }\textbf {\bibinfo {volume} {74}} (\bibinfo {year}
  {2020})}\BibitemShut {NoStop}%
\bibitem [{\citenamefont {Gardiner}\ and\ \citenamefont
  {Collett}(1985)}]{Gardiner1985PRA}%
  \BibitemOpen
  \bibfield  {author} {\bibinfo {author} {\bibfnamefont {C.~W.}\ \bibnamefont
  {Gardiner}}\ and\ \bibinfo {author} {\bibfnamefont {M.~J.}\ \bibnamefont
  {Collett}},\ }\bibfield  {title} {\bibinfo {title} {Input and output in
  damped quantum systems: Quantum stochastic differential equations and the
  master equation},\ }\href {https://doi.org/10.1103/PhysRevA.31.3761}
  {\bibfield  {journal} {\bibinfo  {journal} {Phys. Rev. A}\ }\textbf {\bibinfo
  {volume} {31}},\ \bibinfo {pages} {3761} (\bibinfo {year}
  {1985})}\BibitemShut {NoStop}%
\bibitem [{\citenamefont {Hofer}\ \emph {et~al.}(2011)\citenamefont {Hofer},
  \citenamefont {Wieczorek}, \citenamefont {Aspelmeyer},\ and\ \citenamefont
  {Hammerer}}]{Hofer2011PRA}%
  \BibitemOpen
  \bibfield  {author} {\bibinfo {author} {\bibfnamefont {S.~G.}\ \bibnamefont
  {Hofer}}, \bibinfo {author} {\bibfnamefont {W.}~\bibnamefont {Wieczorek}},
  \bibinfo {author} {\bibfnamefont {M.}~\bibnamefont {Aspelmeyer}},\ and\
  \bibinfo {author} {\bibfnamefont {K.}~\bibnamefont {Hammerer}},\ }\bibfield
  {title} {\bibinfo {title} {Quantum entanglement and teleportation in pulsed
  cavity optomechanics},\ }\href {https://doi.org/10.1103/PhysRevA.84.052327}
  {\bibfield  {journal} {\bibinfo  {journal} {Phys. Rev. A}\ }\textbf {\bibinfo
  {volume} {84}},\ \bibinfo {pages} {052327} (\bibinfo {year}
  {2011})}\BibitemShut {NoStop}%
\end{thebibliography}
\end{document}